\documentclass[onecolumn]{revtex4}
\newcommand{\gtsim}{\mbox{{\raisebox{-0.4ex}{$\stackrel{>}{{\scriptstyle\sim
}}
$}}}}
\newcommand{\ltsim}{\mbox{{\raisebox{-0.4ex}{$\stackrel{<}{{\scriptstyle\sim
}}
$}}}}
\usepackage{graphicx}
\begin{document}
\title{Experimental demonstration of a new radiation mechanism:
emission by an oscillating, accelerated, 
superluminal polarization current}

\author{A. Ardavan$^1$, J. Singleton$^2$, H. Ardavan$^3$,
J.~Fopma$^4$, D. Halliday$^4$ and W. Hayes$^1$}
\affiliation{$^1$Condensed Matter Physics, Department of Physics, 
University of Oxford, The Clarendon Laboratory,
Parks Road, Oxford~OX1~3PU, United Kingdom\\
$^2$National High Magnetic Field Laboratory, TA-35, MS-E536,
Los Alamos National Laboratory, Los Alamos, NM87545, USA\\
$^3$Institute of Astronomy, University of Cambridge,
Madingley Road, Cambridge CB3 0HA, United Kingdom\\
$^4$Central Electronics, Department of Physics,
University of Oxford, The Denys Wilkinson Laboratory,
Keble Road, Oxford~OX1~3RH, United Kingdom 
}
\begin{abstract}
We describe the experimental implementation of a superluminal ({\it
i.e.}\ faster than light {\it in vacuo}) polarization current
distribution that both oscillates and undergoes centripetal
acceleration.  Theoretical treatments predict that the radiation
emitted by each volume element of the superluminally-moving
distribution pattern will comprise a 
\v Cerenkov-like envelope with two sheets that meet along a cusp.
Correspondingly, the emission from the experimental machine is found
to be tightly beamed in both the azimuthal and polar directions.  The
beaming is frequency independent and has a sharply-defined and
unchanging geometry determined only by the speed and path of the
moving distribution pattern, {\it i.e.}\ by the parameters governing
the structure of the \v Cerenkov-like envelopes.  In addition, over a
restricted range of angles, we detect the presence of cusps in the
emitted radiation.  These, which are due to the focusing of wave
fronts on a propagating space curve, result in the reception, during
a short time period, of radiation emitted over a considerably longer
period of (retarded) source time.  The intensity of the radiation at
these angles was observed to decline more slowly with increasing
distance from the source than would the emission from a conventional
antenna.  The angular distribution of the emitted radiation and
the properties associated with the cusps are in good {\it
quantitative} agreement with theoretical models of superluminal
sources once the effect of reflections from the earth's surface are
taken into account.  In particular, the prediction that the beaming
and the slow decay should extend into the far zone has been tested to
several hundred Fresnel distances (Rayleigh ranges).  The excellent
agreement between the theoretical calculations and the data suggests
that the apparatus achieves precise and reproducible control of the
polarization current and that similar machines could be of general
interest for studying and utilizing the novel effects associated
with superluminal electrodynamics.
\end{abstract}

\maketitle

\section{Introduction}
Serious scientific interest in the emission of electromagnetic
radiation by charged particles moving faster than the speed of light
{\it in vacuo} began with the work of Sommerfeld in 1904 and
1905~\cite{HA1}.  The timing of this work was perhaps unfortunate, as
the special theory of relativity was published within a few
months~\cite{einstein}; any known particle that has a charge has also
a rest mass and so cannot be accelerated through the speed of light.
Moreover, an additional conceptual barrier discouraged further work:
even a massless particle cannot move faster than light {\it in vacuo}
if it is charged, because if it did, it would give rise to an
infinitely strong electromagnetic field on the envelope of the wave
fronts that emanate from it~\cite{HA2}.

By and large, this field remained dormant until Ginzburg and
colleagues~\cite{HA2,HA2A,HA2B} pointed out that, though no
superluminal source of electromagnetic fields can be point-like, there
are no physical principles preventing {\it extended} faster-than-light
sources.  The coordinated motion of aggregates of subluminally-moving
charged particles can give rise to macroscopic polarization currents
whose distribution patterns move superluminally~\cite{HA2,HA2A,HA2B}.
Further relevant progress occurred with the theoretical prediction
that extended sources that move faster than their own waves could be
responsible for the extreme properties of both the electromagnetic
emission from pulsars (rapidly spinning, magnetized neutron
stars)~\cite{HA3,HA3A,HA3B,HA3C,pulsar} and the acoustic emission by
supersonic rotors and propellers~\cite{lowson}.  
Finally, theoretical treatments of a particular class of source of
this type~\cite{HAAAJS1,HAAAJS2} 
were stimulated by the design and
construction of the apparatus that is the 
subject of the current paper.
This apparatus implements 
an extended polarization current distribution that has the
time dependence of a traveling wave with centripetally-accelerated
superluminal motion and an oscillating amplitude.
Refs.~\cite{HAAAJS1,HAAAJS2} point out 
several distinctive characteristics of the emission by such
sources:
\begin{enumerate}
\item
The radiation emitted from each volume element of the 
superluminally-moving polarization-current
distribution pattern comprises
a \v Cerenkov-like envelope with 
two sheets that meet along a cusp.
The combined effect of the volume elements leads to a
frequency-independent beaming in both the azimuthal and polar
directions whose geometry is determined by the structure 
of the \v Cerenkov-like 
envelopes~\cite{HA3,HAAAJS1,HAAAJS2}.
\item
The above-mentioned cusps, which spiral upwards and
outwards from the source into the far 
zone~\cite{HA3,HAAAJS1,HAAAJS2},
result from the centripetal acceleration of the source; without
acceleration, the wave fronts would merely pile up on a
two-dimensional envelope~\cite{HA3,lowson}.  The cusps result in the
reception, during a short time period, of radiation emitted over a
considerably longer period of (retarded) source time.
Consequently, the
intensity of the radiation at these angles
declines more
slowly with increasing distance from the source than would the
emission from a conventional antenna.
Note, however, that
energy conservation is not violated; large intensities on the cusp are
compensated by weaker radiation fields elsewhere. 
The propagating cusp
is being reconstructed constantly from conventional 
({\em i.e.}~spherically-decaying) 
waves that combine and then disperse.
\item
Under certain conditions, the emitted radiation contains a broad
spectral distribution of high frequencies that are not present in the
synthesis of the moving, oscillating polarization current.  These
originate from the centripetal acceleration, and from space-time
discontinuities forced on the polarization-current distribution by the
constraints of its circular path.
\end{enumerate}

Based on the above ideas, we have constructed an experimental
realization of an extended, oscillating superluminal source that
undergoes centripetal acceleration.  The present paper describes the
basic principles of the machine, and its use to demonstrate the first
two theoretical findings listed above; the high frequency emission
will be the subject of a future work.  We find that the emission of
the source is well described by the models of
Refs.~\cite{HAAAJS1,HAAAJS2}, confirming the validity of extended
superluminal polarization currents as precisely controllable sources
of electromagnetic radiation with novel properties.

The paper is organized as follows.  Section~\ref{experiment} gives a
description of the experimental considerations necessary to understand
the data from the superluminal source.  Since practical superluminal
sources are relatively novel, and since the raw experimental data are
affected by the proximity of the source to the ground in a way familar
only to communications engineers~\cite{ground}, this section is by
necessity rather detailed; without a thorough explanation of these
effects, the experimental data are not readily understood.  Within it,
Subsection~\ref{expa} describes how the polarization current
distribution is created and animated (set in motion) at superluminal
speeds within a curved strip of alumina. The experimental geometry,
and in particular its relationship to the coordinate systems used in
the theoretical papers~\cite{HA3,HAAAJS1,HAAAJS2} is discussed in
Subsection~\ref{expb}, whilst the detection system and the effects of
interference from the ground are described in Subsection~\ref{expc}.
Experimental results commence in Section~\ref{beams}, which describes
data that conclusively demonstrate the three-dimensional angular
distributions of the \v Cerenkov-like envelopes emitted by the source.
Section~\ref{nonspherical} describes range tests exhibiting the
non-spherically-decaying radiation close to the cusp; in spite of the
non-ideal nature of the experimental machine and the measurement
geometry (which complicated by interference from the ground)
it is possible to
conclusively identify this feature within the emission data.  A
discussion, conclusions and summary are given in
Section~\ref{discussion}. The paper ends with an Appendix describing
the mathematical treatment of interference due to reflections from the
ground. In addition, the Appendix details the fitting methods used to
obtain the exponent of the non-spherical decay associated with the
cusps, and summarizes the parts of the theory given in
Refs.~\cite{HAAAJS1,HAAAJS2} that are used to model the beaming
observed in the experimental data.
\section{Experimental Details}
\label{experiment}
\subsection{Animation of a superluminal polarization current}
\label{expa}
\begin{figure}[tbp]
   \centering
\includegraphics[height=9cm]{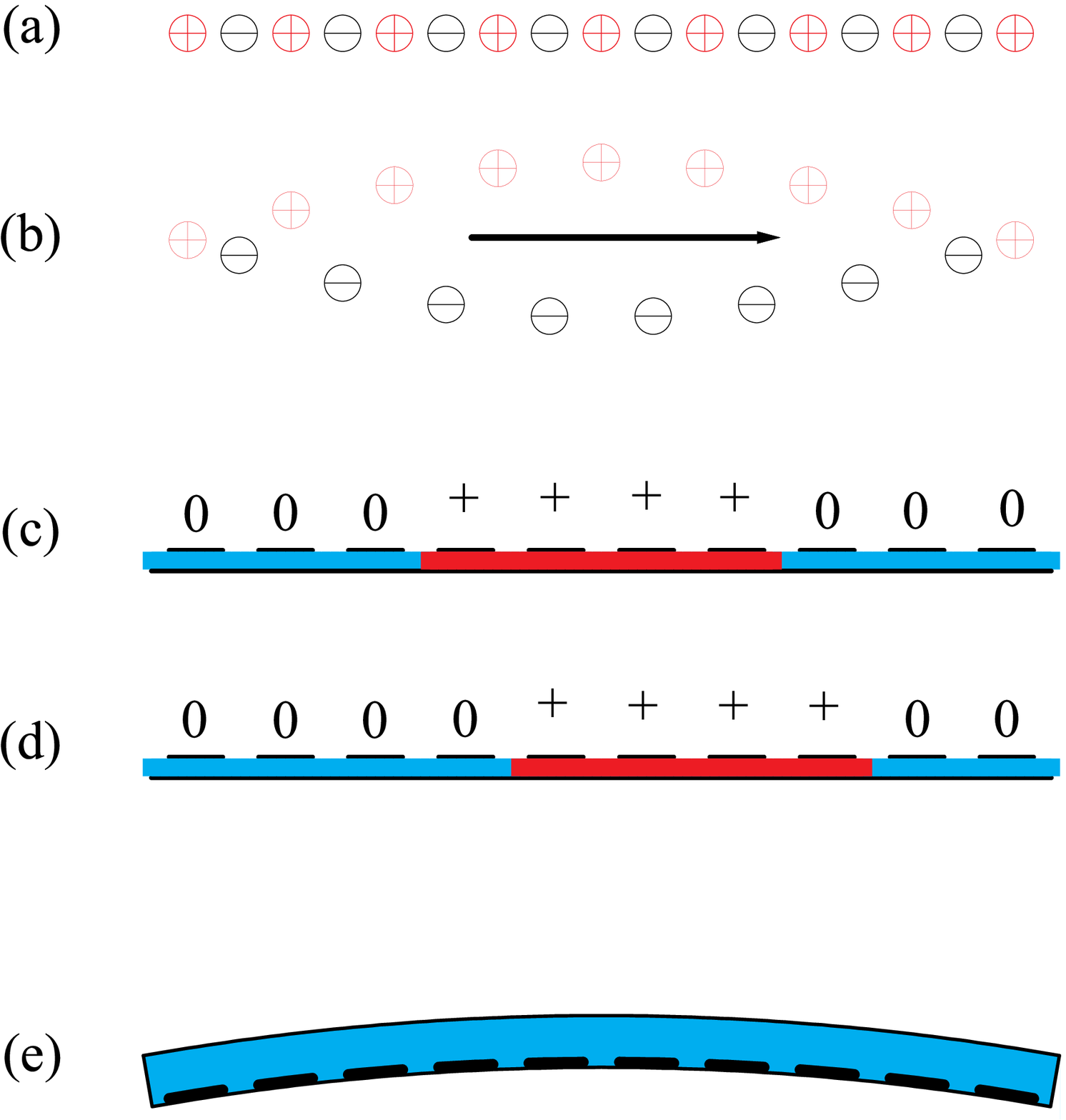}
\includegraphics[height=9cm]{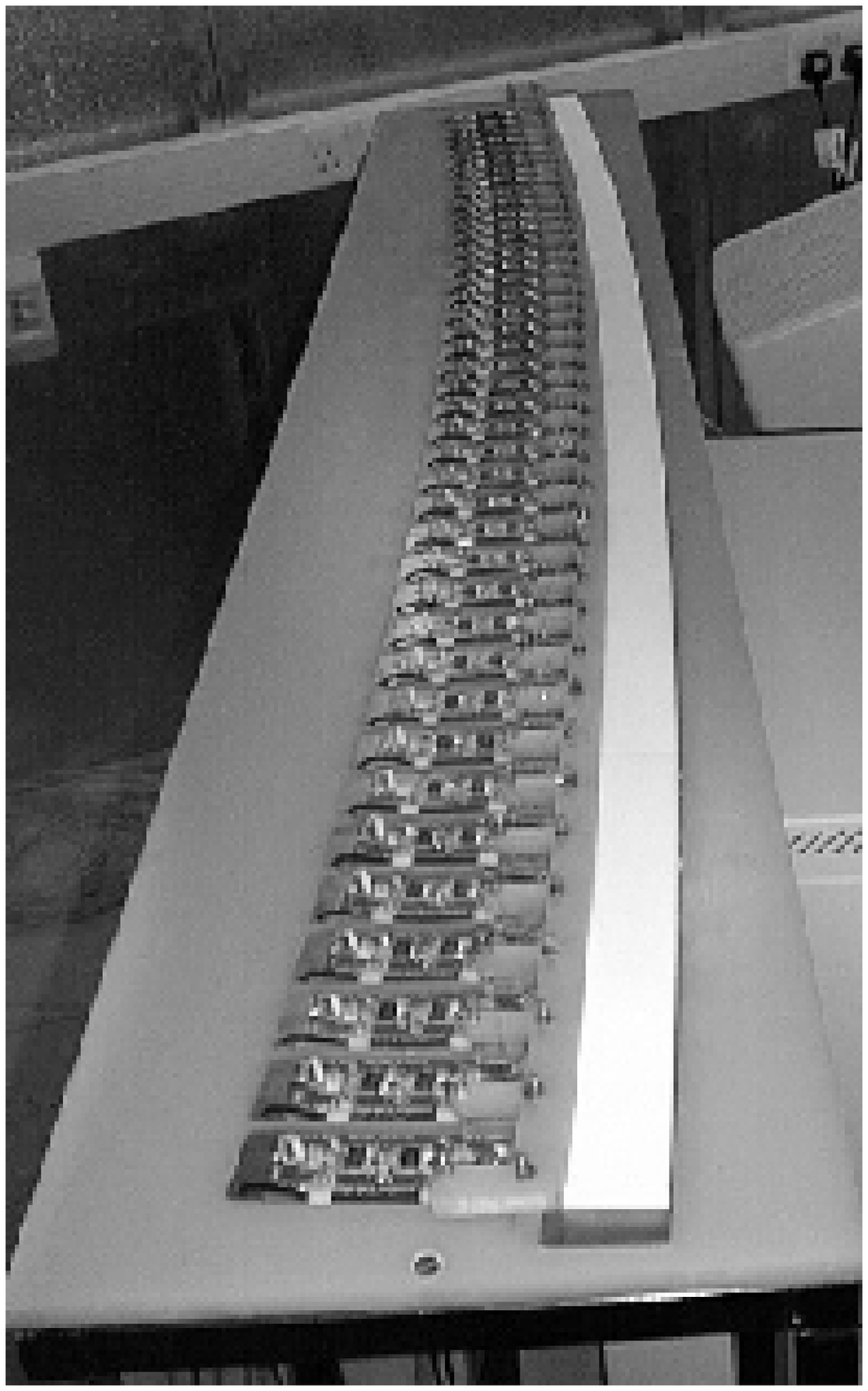}
\caption{Left: experimental animation of a superluminal polarization
current.  (a)~A simplified dielectric solid containing negative
($\ominus$) and positive ($\oplus$) ions. In (b), a spatially-varying
electric field has been applied, causing the positive and negative
ions to move in opposite directions; a finite polarization {\bf P} has
therefore been induced.  If the spatially-varying field is made to
move along the direction of the arrow, the polarized region moves
with it.  (c)~Schematic side view of a practical superluminal source,
showing metal electrodes above a strip of dielectric (shaded region)
and a ground plate below it.  ``0'' indicates that there is no voltage
on that particular upper electrode; the symbol + indicates a positive
voltage applied to the upper electrode.  The voltage on the electrodes
produces a finite polarization of the dielectric (darker shading).
(d)~By switching the voltages on the electrodes on and off, the
polarized region (darker shading) can be made to move along the
dielectric.  (e)~Top view, showing the curvature of the dielectric
(lighter shaded region). The curvature 
introduces centripetal acceleration in
the moving polarized region.  Note also that the electrodes (black
shading) cover only part of the top surface of the dielectric.  In the
discussion that follows, we refer to this section of the
source (electrodes, dielectric and groundplate) as the ``array'', for
convenience.  Right: a photograph of the experimental array;
the copper shields have been removed from the 41 amplifiers
so that their components are visible.  }
\label{fig1}
\end{figure}

The first step in investigating superluminal effects is to identify a
source that can travel faster than the speed of light {\it in vacuo}
without violating Einstein's theory of special
relativity~\cite{einstein}.  The required source explicitly appears in
the Amp\`ere-Maxwell equation, which underlies the emission of
electromagnetic radiation~\cite{bleaney,landau}:
\begin{equation}
\nabla \times {\bf H} = {\bf J}_{\rm free} 
+ \frac{\partial {\bf D}}{\partial t} ={\bf J}_{\rm free} 
+ \epsilon_0\frac{\partial {\bf E}}{\partial t}
+ \frac{\partial {\bf P}}{\partial t}~~~~~{\rm (SI~units)}.
\label{maxwell4}
\end{equation}
Here {\bf H} is the magnetic field strength, ${\bf J}_{\rm free}$ is
the current density of free charges, ${\bf D}$ is the displacement,
${\bf E}$ is the electric field and ${\bf P}$ is the polarization;
the term $\partial {\bf P}/\partial t$ represents the {\it
polarization current density}.  
An oscillating or accelerating
current density ${\bf J}_{\rm free}$ emits electromagnetic radiation
({\em i.e.}\ it acts as a source) and is well-understood; this is the
basis of conventional radio transmission~\cite{bleaney,Stratton} 
and the use of
synchrotrons as light sources~\cite{synchrotron}.  However, as a
current density of free charges is composed of particles such as
electrons or protons, Einstein's theory of special relativity
naturally forbids such a source from traveling faster than the speed
of light.  By contrast, there is no such restriction on a polarization
current.  Fig.~\ref{fig1}(a)-(b) shows a simplified dielectric solid
containing negative and positive ions. In (b), a spatially-varying
electric field has been applied, causing the positive and negative
ions to move in opposite directions.  A finite polarization {\bf P}
has therefore been induced.  If the spatially-varying field is made to
move, the polarized region moves with it; we have a traveling ``wave''
of {\bf P} (and also, by virtue of the time dependence imposed by
movement, a traveling wave of $\partial {\bf P}/\partial t$). Note
that this ``wave'' can move arbitrarily fast 
({\it i.e.} faster than
the speed of light) because the individual ions suffer only small
displacements perpendicular to the direction of the wave and therefore
do not themselves move faster than the speed of light.

\begin{table}[tbp] \centering
\begin{tabular}{|l|l|}
\hline
 Symbol  & Definition \\ \hline
 $\omega$ &  Angular rotation frequency of the distribution 
pattern of the source.\\ \hline
 $\Omega$ &  The angular frequency with which the source oscillates 
(in addition to moving).\\ 
 ~ & One of the two angular frequencies used to drive the experimental 
source (Eq.~\ref{volt1}).\\
\hline
 $f,~f_{\pm}$ &  Frequency of the radiation generated by the source; 
$f_{\pm}=|\Omega \pm m\omega|/2 \pi$. \\ \hline
 $n=2\pi f/\omega$ &   The harmonic number associated with a particular 
radiation frequency.\\ \hline
 $m$ &  The number of cycles of the sinusoidal wave train representing the azimuthal\\
 ~ & dependence of the rotating source distribution [see Eq.\ref{haeq7}] around the \\
 ~& circumference of a circle centred on, and normal to, the rotation axis. \\ \hline
$\eta = m\omega$ & The second of the two angular frequencies used to drive the
experimental source (Eq~\ref{volt1}).\\ \hline
\end{tabular}
\caption{Definition of the various frequencies and numbers
used to describe the source and the emitted radiation, 
after Refs.~\cite{HAAAJS1,HAAAJS2}.}
\label{table1}
\end{table}

Thus far, the bulk of the theoretical analysis of superluminal light
sources has been carried out for rotating, oscillating polarization
current distributions~\cite{HAAAJS1,HAAAJS2}, which possess a
polarization given (in cylindrical polar coordinates) by~\cite{fn1}
\begin{equation}
{\bf P}(r, \varphi, z, t) = {\bf s}(r,z)\cos[m(\varphi-\omega t)]\cos\Omega t.
\label{haeq7}
\end{equation}
Here {\bf s} is a vector field describing the orientation of the
polarization; it vanishes outside the active volume of the
source~\cite{HAAAJS1}.  The other parameters are defined in
Table~\ref{table1}.  The experimental superluminal source described in
this paper represents a discretized version of Eq.~\ref{haeq7}.  It
consists of a continuous strip of alumina ({\em i.e.}\ the material to
be polarized by the applied electric field) on top of which is placed
an array of metal electrodes; underneath is a continuous ground
plate. This forms what is in effect a series of capacitors (a
schematic is shown in Fig.~\ref{fig1}(c)-(e)).  Each upper
``capacitor'' electrode is connected to an individual amplifier;
turning the amplifiers on and off in sequence moves a polarized region
along the dielectric (Fig.~\ref{fig1}(c)-(d)).

In practice, the dielectric in the experimental source is a strip
corresponding to a $10^{\circ}$ arc of a circle of average radius $a=
10.025$~m, made from alumina 10~mm thick and 50~mm across.  Above the
alumina strip, there are 41 upper electrodes of mean width 42.6~mm,
with centers 44.6~mm apart.  Whilst the ground plate is the same width
as the dielectric strip, each electrode covers only the inner 10~mm of
the upper surface (shown schematically in Fig~\ref{fig1}(e)). This has
the effect of producing a polarization in the alumina with both radial
and vertical components (see Eq.~\ref{haeq7}).  In the discussion that
follows, we shall refer to the assembly consisting of groundplate,
dielectric and electrodes as ``the array'' for convenience.

\begin{figure}[tbp]
   \centering
\includegraphics[height=9cm]{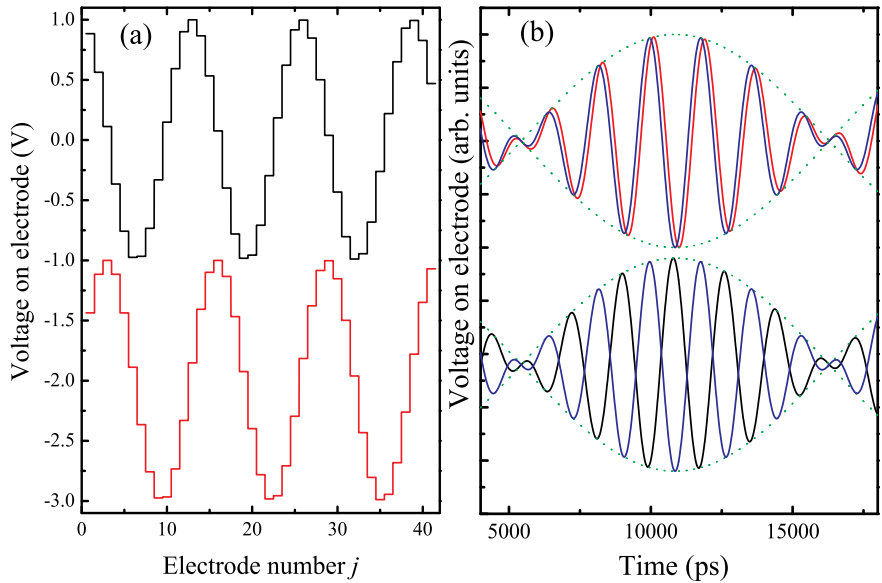}
\caption{(a)~The upper curve shows the instantaneous values of the
voltages on the 41 upper electrodes of the array; $\eta /2\pi =
552.653$~MHz and $\Delta t = 140$~ps, corresponding to $v \approx
1.064 c$ for $a=10.025~$m and $\Delta \varphi = \pi/(18 \times 40)$
(see Eq.~\ref{volt1}).  The resulting polarizations under the
electrodes will be proportional to these voltages.  The lower curve
(offset for clarity) gives the instantaneous voltages 420~ps later;
the polarization ``wave'' has moved along the dielectric.  (Note that
the $\cos \Omega t$ term in Eq.~\ref{volt1} hardly changes during this
time.)  (b)~Example of the time-dependent voltages on array electrodes
20 and 21 (upper curves) and 20 and 26 (lower curves) for the same
parameters as in (a).  Note that the lower-frequency temporal
modulation of each electrode voltage produced by the second cosine
term in Eq.~\ref{volt1} (dotted curves) is in phase for all
electrodes.}
\label{fig2}
\end{figure}

Each upper electrode of the array is driven by an individual shielded
amplifier just behind it; to simulate Eq.~\ref{haeq7}, the $j$th ($j =
1$, 2, 3....) electrode is supplied with a voltage
\begin{equation}
V_j = V_0 \cos[\eta (j\Delta t -t)]\cos \Omega t.
\label{volt1}
\end{equation}
Comparison with Eq.~\ref{haeq7} shows that the angular frequency $\eta
= m \omega$; $\Delta t = \Delta \varphi/\omega$, where $\Delta \varphi
= \pi/(18 \times 40)$ is the angle subtended by the effective center
separation of adjacent electrodes.  The velocity $v$ with which the
polarization current distribution propagates along the dielectric is
set by adjusting $\Delta t$ to give $v = a\Delta \varphi/ \Delta t$
(see Fig.~\ref{fig2}(a)).  Given the dimensions of the experimental
array, $v>c$ is achieved for $\Delta t < 148.8$~ps. (At this point it
should be stated that the modulation scheme used to animate the
superluminal polarization current is quite distinct from any used thus
far in conventional phased-array antennae~\cite{phased}.)

The application of the electrode voltages produces a stepped
approximation to a sinusoidal polarization-current wave that moves at
speed $v$ along the dielectric (see Fig.~\ref{fig2}(a)).  Detailed
analysis of this type of waveform (see Ref.~\cite{HAAAJS1}, Appendix)
shows that the dominant emission is almost identical to that produced
by a pure sinusoidal wave with the same wavelength and amplitude.
Most of the emission occurs at two
frequencies, $f_{\pm} = |\Omega \pm \eta|/2\pi \equiv |\Omega \pm m
\omega|/2\pi$, as indicated in Table~\ref{table1}.

As mentioned above, the speed of the source ({\em i.e.} the speed at
which the sinusoidal polarization current ``wave'' propagates along
the dielectric) is set by the parameter $\Delta t$.  Experimentally,
this is done by observing the time dependence of the voltages on each
electrode (Figure~\ref{fig2}(b)) using a high-speed oscilloscope.

Finally, it should be emphasised that great care was taken to
minimize unwanted radiation from the amplifiers, cabling
and signal generators of the apparatus to ensure that
the detected signal was dominated by the emission
from the polarization current. Steps taken included
proper encasing of all of the signal generation equipment,
the enclosure of the 41 amplifiers inside thick copper shields
(not shown in Fig.~\ref{fig1}) and the correct impedance matching
of connections. 

\subsection{The experimental geometry}
\label{expb}
To map out the spatial distribution of the radiation from the
superluminal source, one would ideally position the array in empty
space and move a detector to points around it, covering a
comprehensive range of angles and distances.  However, practical
experiments must be conducted close to the earth's surface, where
reflections from the ground are a complicating factor~\cite{ground}
(see Appendix); numerous tests showed that highly reproducible data
are obtained when the ``ground'' under the path between array and
detector is smooth, level and of a uniform quality~\cite{fn2}.  After
trials at several locations, the bituminised macadam surface of an
airfield runway (Turweston Aerodrome, near Northampton, UK) was found
to be almost ideal in providing a substantial length of
well-characterized surface. All of the experiments
reported in the current paper employ the runway,
which is approximately 900~m long and 20~m wide. A change in
slope about two-thirds of the way along the
runway restricts the range tests reported in 
Section~\ref{nonspherical} to distances of 600~m or less.
However, measurements of the angular distribution
of the radiated power from the array (Section~\ref{beams})
employed the full length of 900~m.

The geometry of the runway 
dictates that the best configuration for
measuring the spatial and angular distribution of the radiation is to
fix the position of the detector (on the runway, at a distance $R$
from the source) and to change the orientation of the array to access
a full range of polar and azimuthal angles.  This is accomplished by
placing the array on a platform that tilts about a horizontal-axis
pivot at one end; this is in turn fixed on a turntable that can rotate
about a vertical axis.

\begin{figure}[tbp]
   \centering
\includegraphics[height=7.5cm]{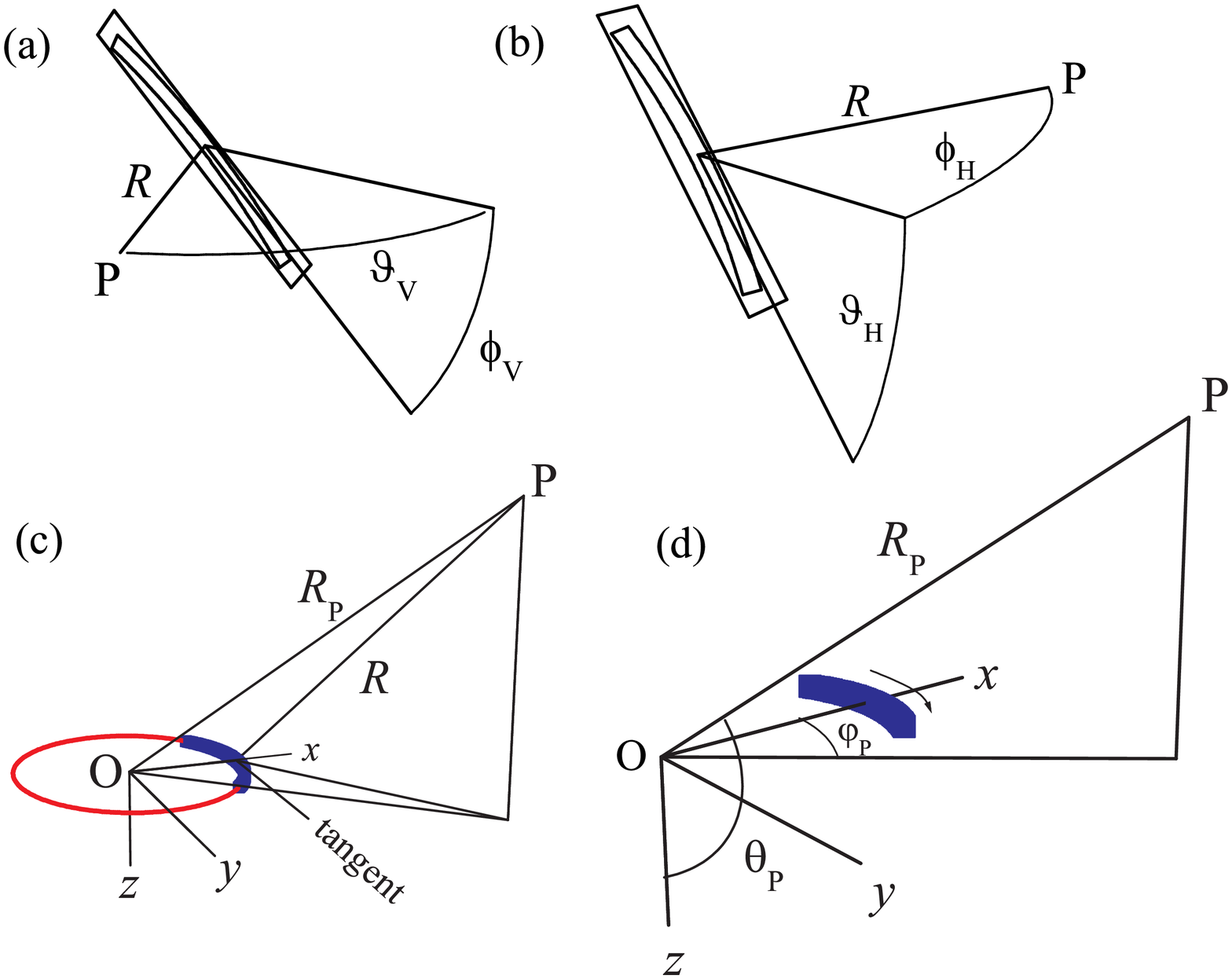}
\includegraphics[height=7.5cm]{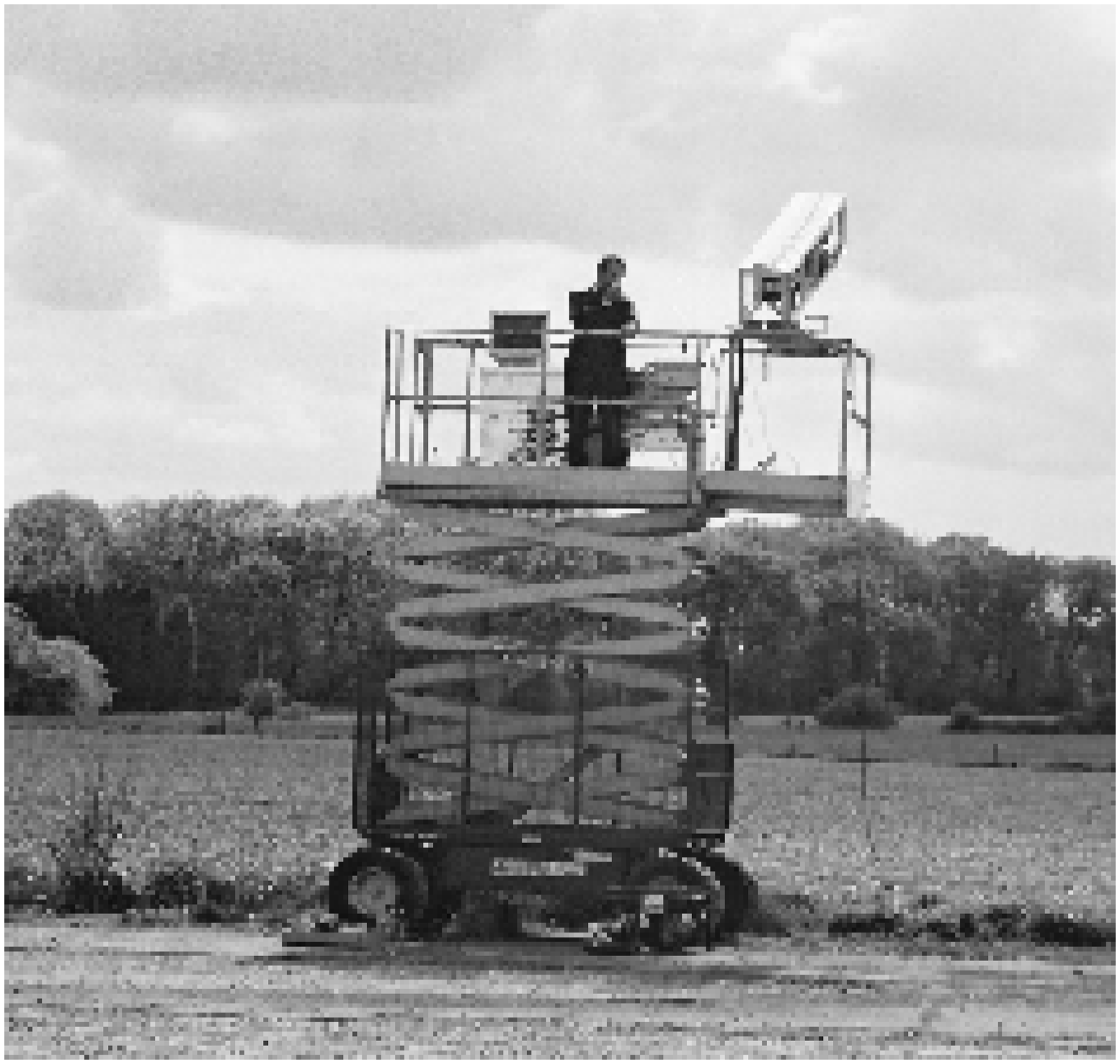}
\caption{Left: the coordinates used to measure the spatial
distribution of radiation from the array.  In the experiments
((a),(b)), the detector is at a fixed point P, distance $R$ from the
array; the array is placed on a platform tilting about a horizontal
axis pivot that is in turn fixed to a turntable rotating about a
vertical axis.  In (a), the array plane is vertical; its long axis is
tilted to an angle $\phi_{\rm V}$ with respect to horizontal. The
turntable is rotated about a vertical axis so that the array plane
makes an angle $\vartheta_{\rm V}$ (in the horizontal plane) with the
line connecting P and the array center.  In (b), the array plane is
initially horizontal; the array plane is then tilted to an angle
$\vartheta_{\rm H}$ with respect to horizontal.  The turntable is
rotated about the vertical axis so that the projection of the long
axis of the array on the horizontal plane makes an angle $\phi_{\rm
H}$ with the line connecting P and the array center. Note that the
coordinate systems are not equivalent; $\tan\phi_{\rm V} = \tan
\phi_{\rm H}/\cos \vartheta_{\rm H}$ and $\sin\vartheta_{\rm V}=
\sin\vartheta_{\rm H}\cos\phi_{\rm H}$.  (c)~The array and detector
positions in terms of the coordinate system used in the Appendix and
Refs.~\cite{HAAAJS1,HAAAJS2}.  In the model, each volume element of
the sourse moves on a circular path centered at the origin O; a
typical path is shown, with the array being represented by thicker
shading.  The array center is at $(a,0,0)$, and the tangent to the
array is parallel to the $y$ axis; the distances, from O, to the
detector ($R_{\rm P}$) and to the array center ($R$) are compared.
(d)~Definition of the spherical polar coordinates $R_{\rm P},
\theta_{\rm P},\varphi_{\rm P}$ used in the Appendix. The arrow
indicates the direction of propagation of the polarization current
distributions in the experimental array.  Right: the array in use
during the experiments.  }
\label{fig3}
\end{figure}

It is possible to mount the array on the platform in two
configurations.  In the first, the array plane is vertical
(Fig.~\ref{fig3}(a)), so that its long axis can be tilted (about the
horizontal pivot) to an angle $\phi_{\rm V}$ with respect to
horizontal; the turntable is then rotated about a vertical axis so
that the array plane makes an angle $\vartheta_{\rm V}$ (in the
horizontal plane) with the line connecting the detector at P and the
array center. In the second configuration, the array plane is
initially horizontal (Fig.~\ref{fig3}(b)); the array plane is then
tilted using the horizontal pivot to an angle $\vartheta_{\rm H}$ with
respect to horizontal.  The turntable is then rotated about a vertical
axis so that the projection of the long axis of the array on the
horizontal plane makes an angle $\phi_{\rm H}$ with the line
connecting the detector at P and the array center.

Finally, to aid comparison of experiment and theory, it is necessary
to relate the experimental spherical polar coordinates
$(R,\vartheta_{\rm V},\phi_{\rm V})$ and $(R,\vartheta_{\rm
H},\phi_{\rm H})$, which are measured relative to the array center, to
the cylindrical and spherical polar coordinates used to evaluate the
integrals in the Appendix and in Refs.~\cite{HAAAJS1,HAAAJS2}.  In the
latter papers, each volume element of the superluminal source
distribution traverses a complete circular orbit centered at the
origin of the coordinate system, and the radial distances to the
individual source elements and to the observation point at $(R_{\rm
P}, \theta_{\rm P},\varphi_{\rm P})$ are specified with respect to the
common center of these circular orbits (see Fig.~\ref{fig3}(c) and
(d)).  Although the experimental source does {\it not} encompass a
complete circle, the high degree of symmetry represented by this
coordinate system greatly facilitates the computation of the integrals
in the Appendix.  The relationship between the experimental
coordinates and the coordinate system used in the Appendix is shown in
Figs.~\ref{fig3}(c) and (d).

\subsection{Detector system, array and detector heights, and interference
from ground reflections}
\label{expc}
The detector was a Schwarzbeck Mess-Elektronik UHA~9105 half-wave
tuned dipole antenna, optimized for the frequency range 300~MHz to
1~GHz.  and mounted on a mobile wooden tower that allowed it to be
fixed at varying heights $h_{\rm D}$ above the ground.  The antenna
was fitted with an internal coaxial BAL-UN to ensure good impedance
matching with an external load and was connected via a 6~m long
impedance-matched coaxial cable ($50~\Omega$) to an Anritsu MS2711B
portable spectrum analyser.  The background noise floor with this
equipment was between -120 and -130~dBm, depending on weather
conditions at the airfield.

In the current paper, the settings of the experimental machine are
such that the emitted radiation is polarized chiefly in directions
parallel to the plane of the array.  When the array is orientated as
in Fig.~\ref{fig3}(a), the detector aerial (see below) is arranged to
detect vertically-polarized radiation; if the array is orientated as
in Fig.~\ref{fig3}(b), the detector is configured to detect
horizontally-polarized radiation.

For the measurements described in the following sections, the array
turntable was mounted on a scissor lift that allowed the height of the
center of the array above the runway to be varied between $h_{\rm A} =
3.3$~m and 5.0~m; for each measurement, the heights $h_{\rm A}$ and
$h_{\rm D}$ were measured to an accuracy of $\pm 5$~cm.  An accurate
measurement of $h_{\rm A}$ and $h_{\rm D}$ is important because the
intensity received at the detector results from interference between
signals traveling along a direct path between array and detector and
reflections from the ground~\cite{ground}. The situation is shown in
Fig.~\ref{intscheme}(a). This process is covered in detail in the
Appendix.

The effect of the interference on the signal received from a simple
horizontally-polarized dipole aerial is shown in
Fig~\ref{intscheme}(b).  Firstly, clear fringes are seen in the
detected intensity as a function of distance.  A second consequence of
the interference is less obvious.  Fig.~\ref{intscheme} shows received
power in the logarithmic units Belm versus $\log_{10}$(distance); the
gradient of the plot should be $-\gamma$, where (received
power)~$\propto$ (distance)$^{-\gamma}$.  Naive expectations based on
the emission of a dipole surrounded by infinite empty space lead one
to predict $\gamma = 2$ (the inverse square law~\cite{bleaney}); in
practice, the interference leads to $\gamma \approx 4$ (see
Fig.~\ref{intscheme}(b)).  Such a dependence is familiar to radio
engineers as the penultimate term of the ``Egli''
relationship~\cite{egli} describing the power lost for frequencies $f$
in the hundreds of MHz range when transmitting over a path close to
the earth's surface:
\begin{equation}
{\rm path~loss~(dB)} = 113+20\log_{10}f+40 \log_{10}
({\rm path~length~(miles)})-20\log_{10}(h_{\rm A}h_{\rm D}).
\label{eglilaw}
\end{equation}
The effects of interference obviously complicate the search for the
non-spherical decay associated with the cusp (see
Section~\ref{nonspherical}). One criterion used in the experiments was
to look for significantly smaller path losses than Egli
(Eq.~\ref{eglilaw}); more detailed analysis was then used to model the
detailed behavior of the data.

Reflections from the ground complicate also the modeling of the
angular distribution of the emission from the array.  A detailed
treatment of the coordinates needed for delineating the direct and
reflected rays is given in the Appendix, as are the equations from
Refs.~\cite{HAAAJS1,HAAAJS2} required for modeling the data.

\begin{figure}[tbp]
   \centering
\includegraphics[height=6cm]{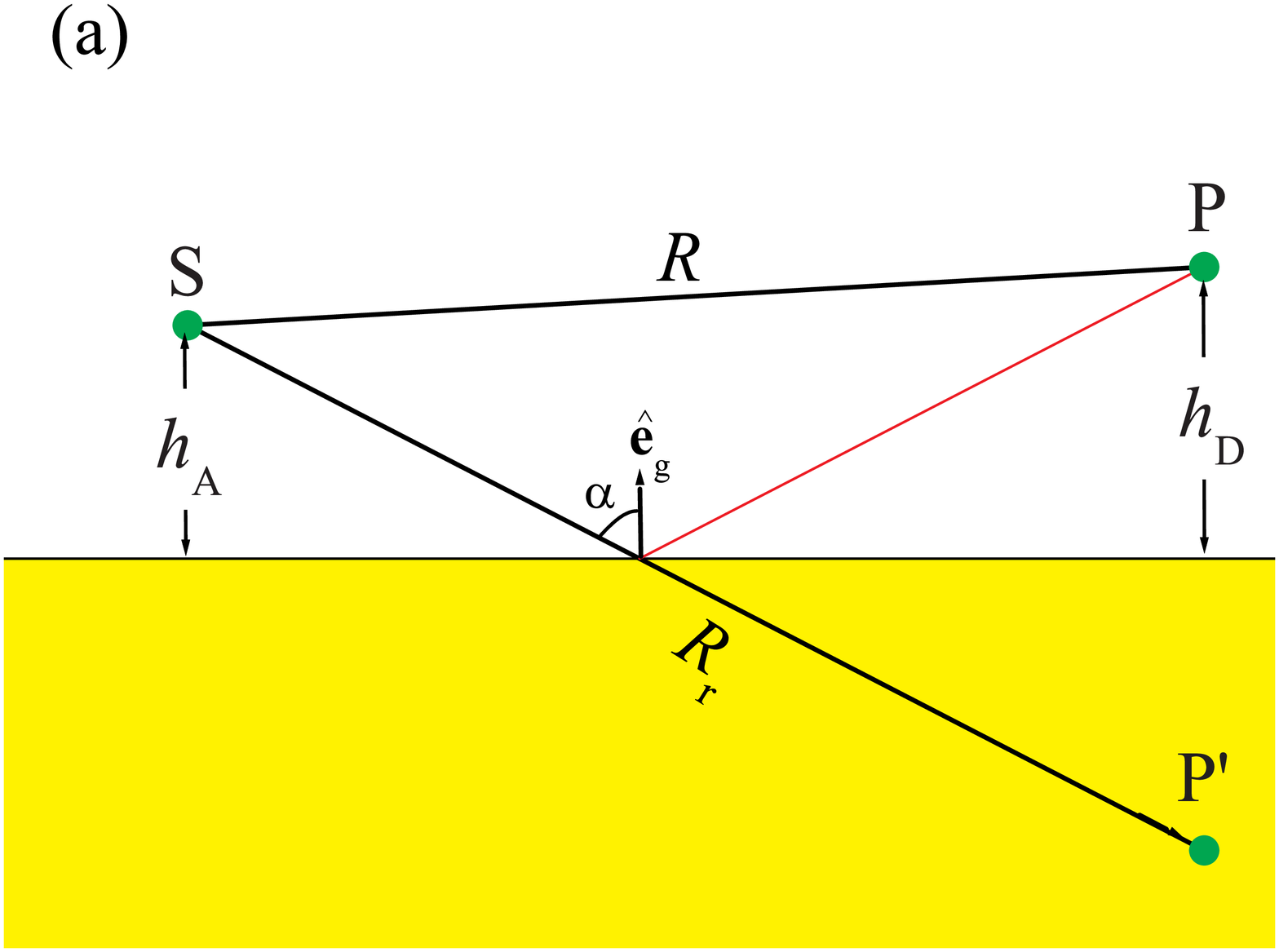}
\includegraphics[height=6cm]{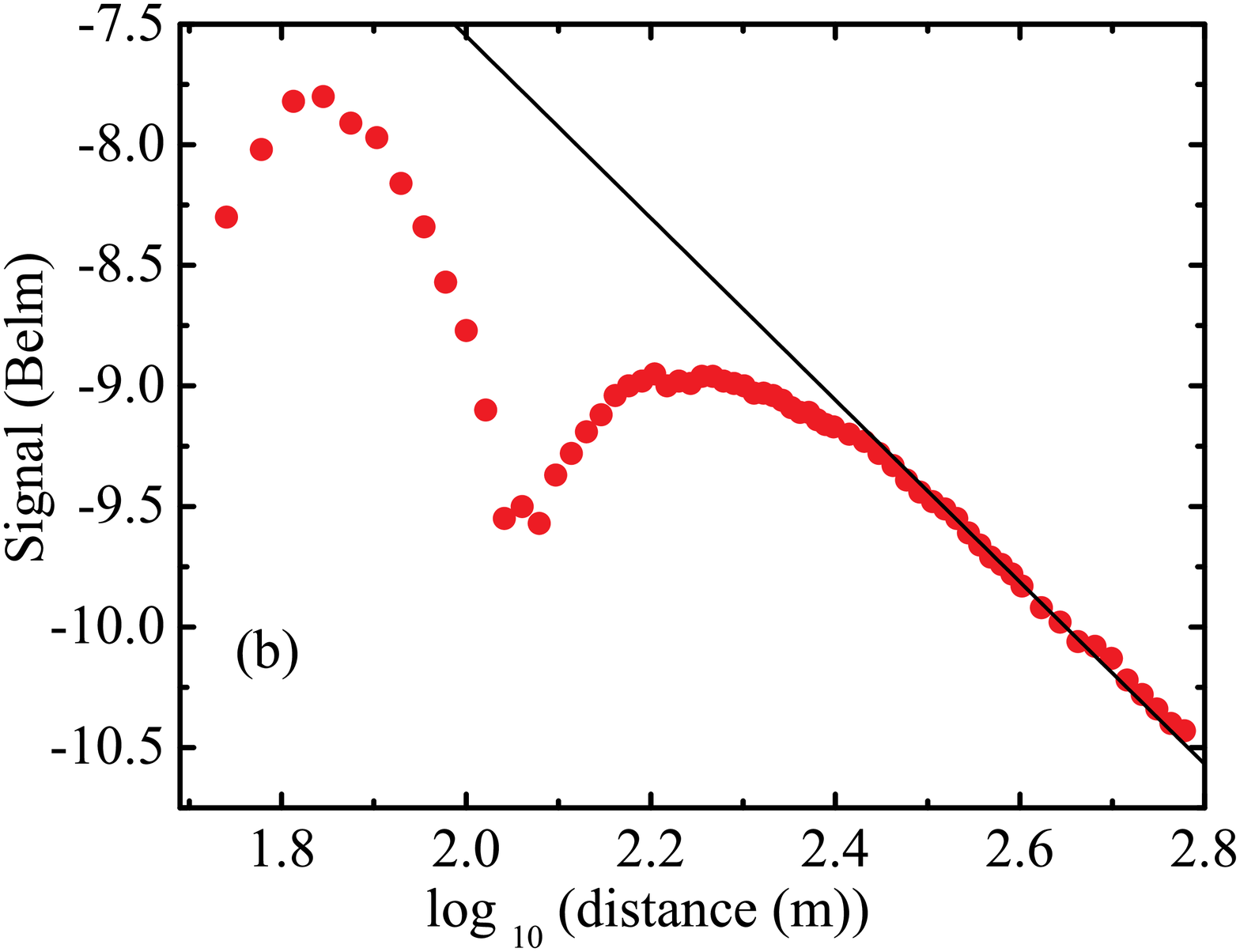}
\caption{(a)~Geometry of the interference between a direct wave from
the source S, height $h_{\rm A}$ above the ground, and a wave
reflected from the ground.  The detector, at height $h_{\rm D}$ above
the ground, is labeled by P. The direct waves travel a path length
denoted by $R$.  The path source-ground-detector is equal in length to
a direct path from the source to an image of P at P${}^\prime$. The
length of this equivalent path is denoted by $R_{\rm r}$.  The other
symbols shown are used in the Appendix.  (b)~Effect of interference on
the power received from a horizontally-polarized dipole aerial excited
at $598.696$~MHz as a function of distance ($h_{\rm A} = 5.12$~m;
$h_{\rm D} = 5.0$~m). The points are data, and the straight line is a
least-squares fit to the data beyond the final interference fringe;
the gradient is $\gamma = -3.8$.}
\label{intscheme}
\end{figure}

\section{Experimental results: intensity distribution of the radiation}
\label{beams}
\subsection{Intensity as a function of $\vartheta_{\rm V}$
at fixed $\phi_{\rm V}$} The array was mounted in the configuration
shown in Fig.~\ref{fig3}(a) and the detector antenna placed at a fixed
distance $R$ away; detector and array were fixed at heights $h_{\rm
D}$ and $h_{\rm A}$ respectively above the level runway.  The angle
$\phi_{\rm V}$ was set to a series of fixed values with an accuracy of
$\pm 0.5^{\circ}$ using a Suunto clinometer, and then the angle
$\vartheta_{\rm V}$ was varied using the turntable. The measurement of
$\vartheta_{\rm V}$ was made to an accuracy of $\pm 0.5^{\circ}$
using a graticule on the turntable.  At each value of ($\vartheta_{\rm
V},\phi_{\rm V}$), the intensity of the signal at the detector antenna
was measured using the spectrum analyser.

\begin{figure}[tbp]
   \centering
\includegraphics[height=10cm]{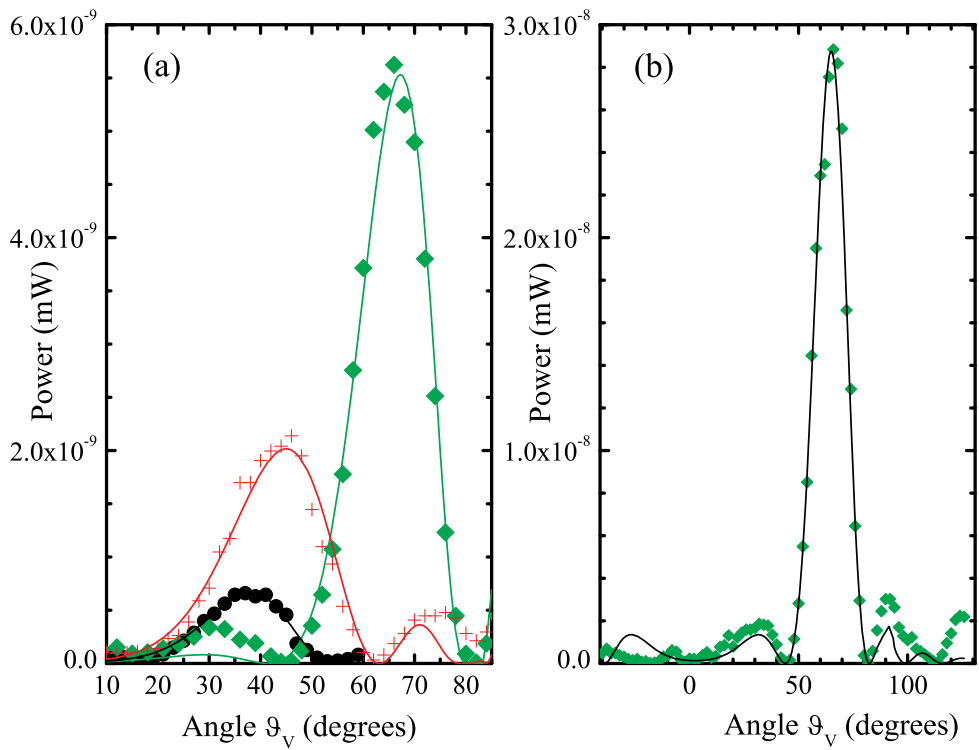}
\caption{(a)~The effect of varying the source speed, $v = a \Delta
\varphi/\Delta t$ on the angular distribution of the radiation from
the array.  For each data set, $\phi_{\rm V} = 2 ^{\circ}$ and the
array to detector distance is $R=600$~m.  Data are points, identified
as follows: filled circles: $v=1.064 c$, $h_{\rm A} = 3.7$~m, $h_{\rm
D} = 4.1$~m; crosses: $v=1.25 c$, $h_{\rm A} = h_{\rm D} = 5.0$~m;
diamonds: $v=2.0 c$, $h_{\rm A} = h_{\rm D} = 5.0$~m.  The curves are
fits, using the model (Eq.~\ref{power}) described in the Appendix.
For each data set, $\eta/2 \pi = 552.654$~MHz, $\Omega/2\pi =
46.042$~MHz, and the observation is made at the higher of the two
emitted frequencies, $f_{+}=|(\Omega +m\omega)|/2\pi \equiv(\Omega +
\eta)/2\pi = 598.696$~MHz.  Note that the peak emission moves out to
higher values of $\vartheta_{\rm V}$ as the source speed increases.
(b)~Power detected versus $\vartheta_{\rm V}$ over a wider angular
range than (a); here, $R=400$~m, $h_{\rm A}=h_{\rm D}=5.0$~m,
$\phi_{\rm V}=2^{\circ}$ and the frequency is the same as in (a). The
points (hollow diamonds) are data; the model (using the experimental
source speed) is the continuous curve.  }
\label{expfiga}
\end{figure}

Fig.~\ref{expfiga}(a) shows the effect of varying the source speed $v
= a\Delta \varphi/\Delta t$ on the intensity distribution as a
function of $\vartheta_{\rm V}$.  Data for $v/c=1.064$, 1.25 and 2.0
are shown; for each data set, $\eta/2 \pi = 552.654$~MHz, $\Omega/2\pi
= 46.042$~MHz, and the observation is made at the higher of the two
emitted frequencies, $f_{+}=|(\Omega +m\omega)|/2\pi \equiv(\Omega +
\eta)/2\pi = 598.696$~MHz.  As the source speed increases, the peak in
detected power moves out to higher values of $\vartheta_{\rm V}$, in
line with expectations; the contributions that are received from
different source elements, by an observer in the far field, are in
phase when (see Appendix, Eqs.~\ref{transform1a} and \ref{eqnf})
\begin{equation}
\theta_{\rm P} =\arcsin\Big({{m c}\over{nv}}\Big),\quad{\rm or}\quad
\vartheta_{\rm V}=\arcsin\Big\{{R\over R_{\rm P}}
\Big[1-\Big({{m c}\over{nv}}\Big)^2\Big]^{1\over2}\Big\}.
\label{emissionangle}
\end{equation}
Here $n=2\pi f/\omega$ is the harmonic number of the emitted radiation
with $f = f_{+}$ or $f_{-}$, and $m=\eta/\omega$ is determined by the
angular frequency $\omega=v/a=\Delta \varphi/\Delta t$ of the rotation
of the source (see Table~\ref{table1})~\cite{HAAAJS1}.  Note that the
data are well reproduced by the model (see Appendix, Eq.~\ref{power})
using the experimental source speeds.

Compared to the emission of a conventional antenna~\cite{vivaldi}, the
rather small side lobes are 
a further characteristic of the beaming
produced by an accelerated superluminal source.  To show this more
clearly, a single data set, corresponding to $v/c = 2.0$, is shown in
Fig.~\ref{expfiga}(b). Although the projection of the length of the
array onto a plane normal to the observation direction is comparable
to or smaller than the radiation wavelength {\it and} the observation
is made at many hundreds of Fresnel distances, most of the power is
transmitted in a beam around $20^{\circ}$ wide, with small side lobes.
This beam structure is reproduced very well by the model (Appendix,
Eq.~\ref{power}), using the experimental source speed as an input
parameter. (In the case of the present emission mechanism, the side
lobes are caused mainly by interference from the ground; it follows
from Eq.~\ref{power} that in free space they would be much weaker.)

\begin{figure}[tbp]
   \centering
\includegraphics[height=10cm]{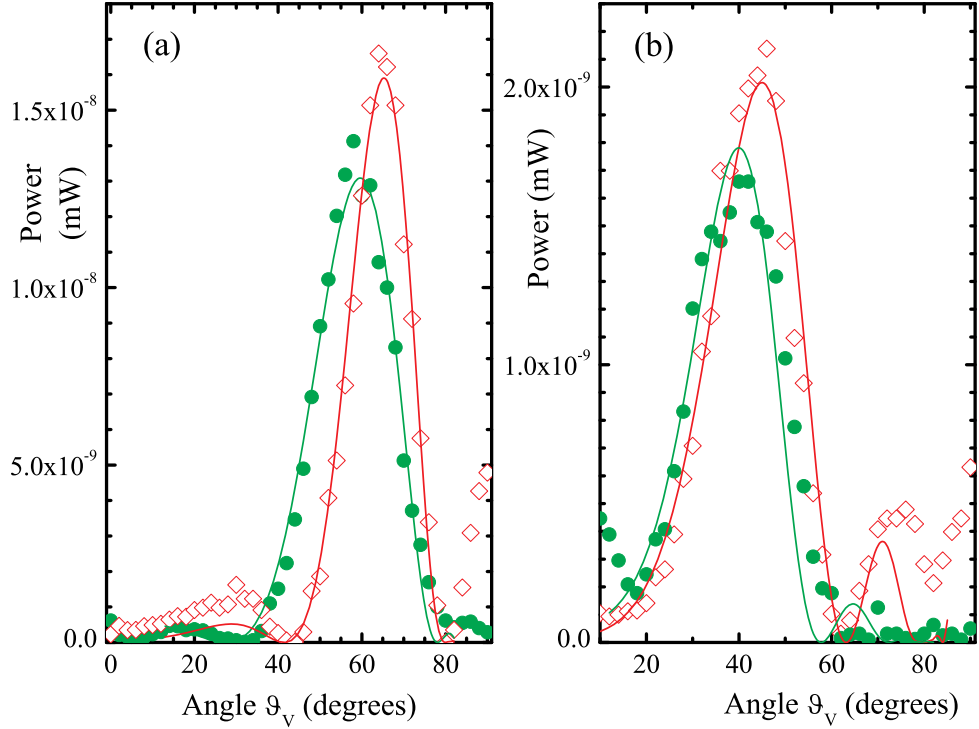}
\caption{Comparison of the emission at $f_{+} = (\eta + \Omega)/2 \pi
=598.696$~MHz (hollow diamonds) or $f_{-} = (\eta - \Omega)/2 \pi
=506.614$~MHz (filled circles) for two different source speeds.  In
(a), $v=2.0c$, $R=400$~m, $h_{\rm A}=3.32$~m, $h_{\rm D}=3.71$~m and
$\phi_{\rm V}=2.5^{\circ}$.  In (b), $v=1.25c$, $h_{\rm A} = h_{\rm D}
= 5.0$~m, $\phi_{\rm V} = 2^{\circ}$ and $R= 600$~m. For all data
sets, the values of $\Omega$ and $\eta$ are the same as those in
Fig.~\ref{expfiga}, and in all cases the curves are generated by the
model (Eq.~\ref{power}) using the experimental values of $f_{\pm}$ and
$v/c$. Note that, in agreement with Eq.~\ref{emissionangle}, the
emission is peaked at higher values of $\vartheta_{\rm V}$ for $f_{+}$
than for $f_{-}$.  }
\label{fcomp}
\end{figure}

Whilst Fig.~\ref{expfiga} shows data for the emission at $f_{+}=
|(\Omega + \eta)|/2\pi$, the description of the source,
Eq.~\ref{haeq7}, leads one to expect simultaneous emission at a second
frequency, $f_{-} = |(\Omega - \eta)|/2\pi$.  As $f_{+}$ and $f_{-}$
have different harmonic numbers, $n$, their angle of emission will
differ slightly (see Eq.~\ref{emissionangle} and Table~\ref{table1}).
This effect is shown for source speeds of $v/c=2.0$ in
Fig.~\ref{fcomp}(a) and $v/c=1.25$ in Fig.~\ref{fcomp}(b).  In each
case, the emission at $f_{+}$ is peaked at a higher value of
$\vartheta_{\rm V}$ than that at $f_{-}$, as predicted by
Eq.~\ref{emissionangle} and the solid curves (see Appendix).

\begin{figure}[tbp]
   \centering
\includegraphics[height=8cm]{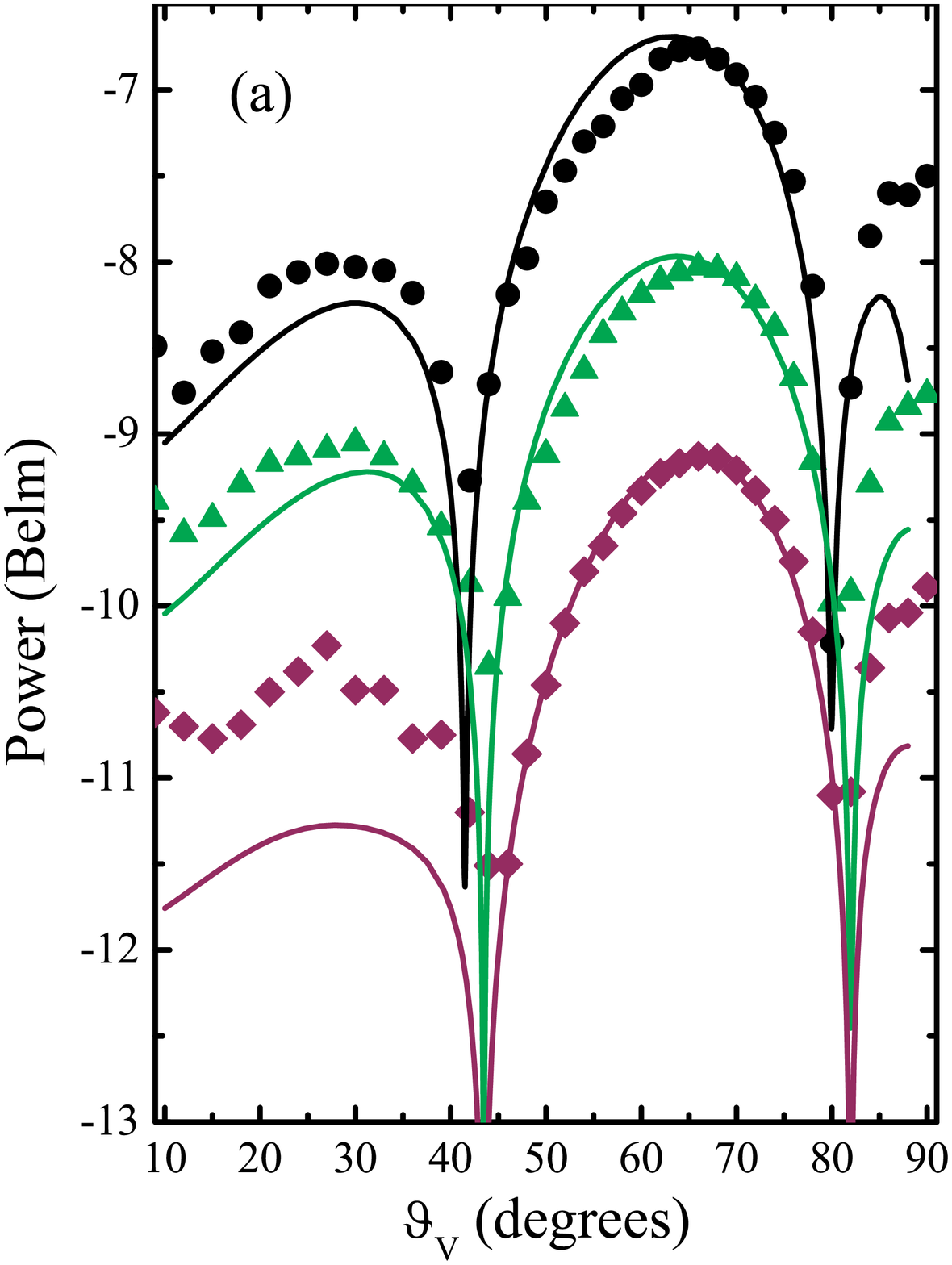}
\includegraphics[height=8cm]{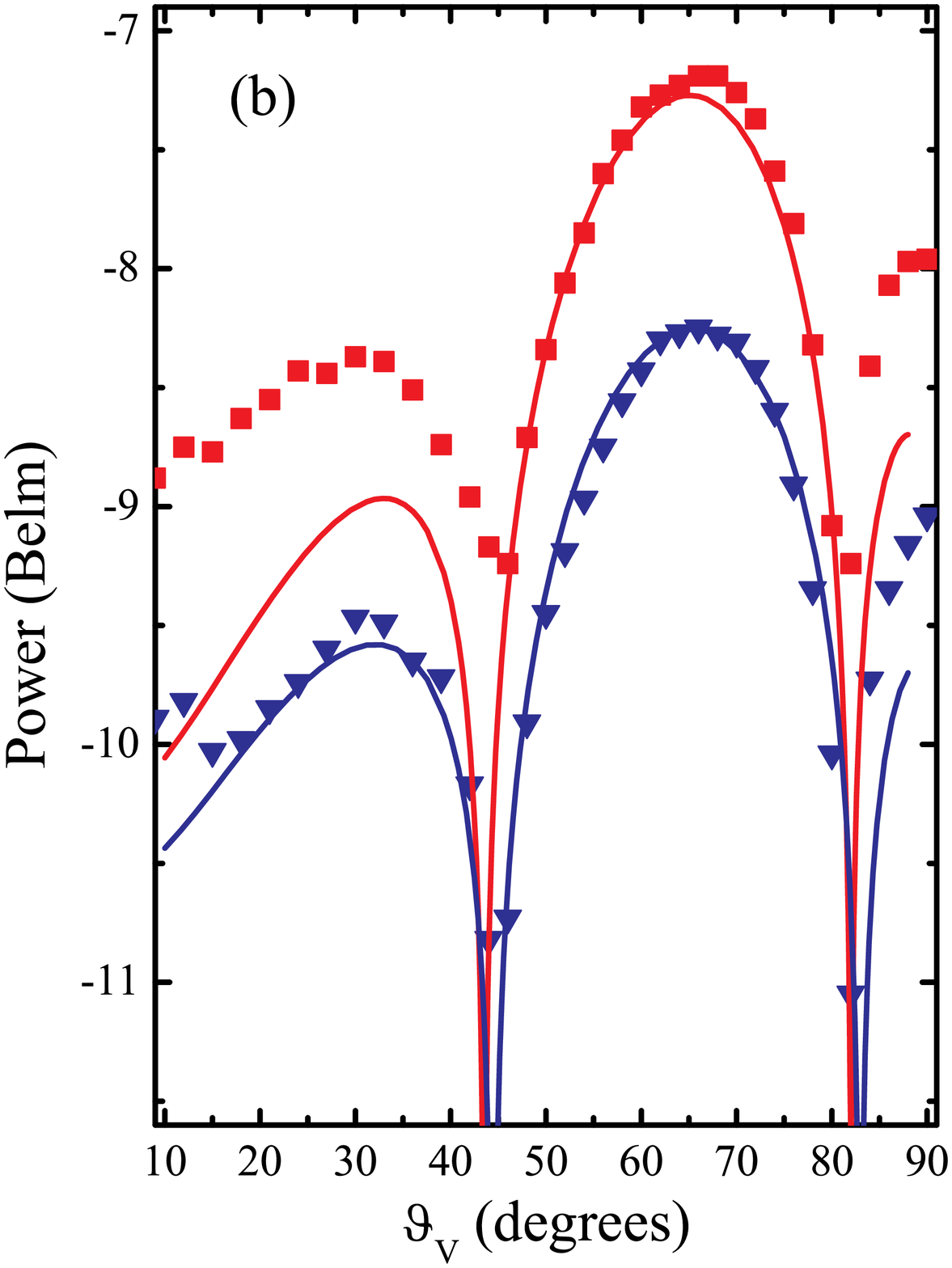}
\caption{(a)~The $\vartheta_{\rm V}$ dependence of the detected power
for $R = 200~{\rm m}$ (circles), $500~$m (triangles), and $900~$m
(diamonds).  (b)~The $\vartheta_{\rm V}$ dependence of the detected
power for $R=300~$m (squares) and $600~$m (inverted triangles).  In
both cases, $f_{+} = (\eta + \Omega)/2 \pi =598.696$~MHz, $v=2.0 c$,
$h_{\rm A} = h_{\rm D} = 5.0$~m and $\phi_{\rm V} = 2^{\circ}$; points
are data and curves are fits to the model in the Appendix
(Eq.~\ref{power}). }
\label{beam2c}
\end{figure}

It has already been mentioned in the discussion of
Fig.~\ref{expfiga}(a) that the beam is rather sharply defined as a
function of $\vartheta_{\rm V}$, even though the projection of the
length of the array on the plane perpendicular to the observation
direction is comparable to or smaller than the radiation wavelength.
This characteristic is independent of distance over the range
measured, as shown in Fig.~\ref{beam2c}, which plots data (points) for
$200~{\rm m}\leq R \leq 900$~m.  For all of these $R$ values, each
representing hundreds of Fresnel distances, the central part of the
beam preserves its overall width and general shape.  This development
of beam structure with $R$ is reproduced by the theoretical
description (Fig.~\ref{beam2c}, curves; see Appendix,
Eq.~\ref{power}), using the experimental source speed as input
parameter.

Note that all of the data sets plotted in this Section involve small
values of $\phi_{\rm V}(\sim 2^{\circ})$.  The reason for this
constraint is that as well as having a peaked structure as a function
of the polar angle $\vartheta_{\rm V}$, the emission of the source is
also peaked in the plane of rotation ({\it i.e.} around $\phi_{\rm
V}=\phi_{\rm H} =0$), as expected in the theoretical
description~\cite{HAAAJS1,HAAAJS2}.  This observation will be
described in the following Section.
\subsection{Intensity as a function of $\phi_{\rm H}$ at fixed 
$\vartheta_{\rm H}$}
\label{phivary}
The array was mounted in the configuration shown in Fig.~\ref{fig3}(b)
and the detector was placed at a fixed distance $R$ away; detector and
array were fixed at the respective heights of $h_{\rm D}$ and $h_{\rm
A}$ above the level runway.  The angle $\vartheta_{\rm H}$ was set to
a series of fixed values to an accuracy of $\pm 0.5^{\circ}$ using a
Suunto clinometer, and then the angle $\phi_{\rm H}$ was varied using
the turntable.  The measurement of $\phi_{\rm H}$ was carried out
using a graticule on the turntable to an accuracy of $\pm
0.5^{\circ}$.  At each value of ($\vartheta_{\rm H},\phi_{\rm H}$)
the intensity was recorded.
\begin{figure}[tbp]
   \centering
\includegraphics[height=9cm]{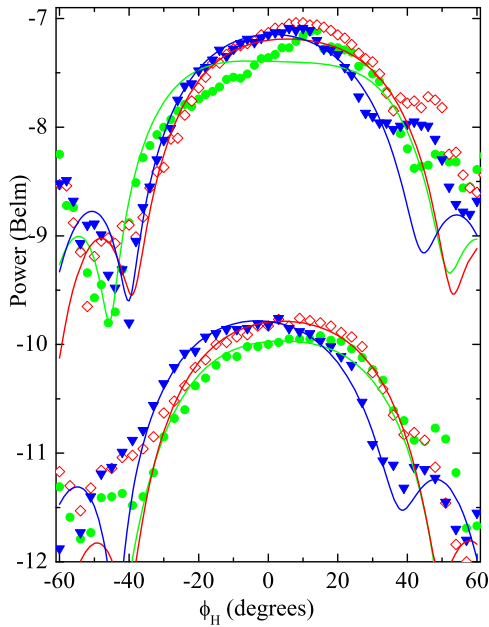}
\caption{The $\phi_{\rm H}$ dependence of the detected power
(logarithmic scale) for $R = 200~{\rm m}$ (upper) and 900~m (lower)
for $\vartheta_{\rm H}=30^{\circ}$ (filled circles), $35^{\circ}$
(hollow diamonds) and $40^{\circ}$ (inverted triangles). Solid curves
are fits to the model in the Appendix (Eq.~\ref{power}). The frequency
is $f_{+} = (\eta + \Omega)/2 \pi =598.696$~MHz, $v=1.064
c$, $h_{\rm A} = h_{\rm D} = 5.0$~m.}
\label{phidata}
\end{figure}

Fig.~\ref{phidata} shows the beam profile as a function of $\phi_{\rm
H}$ for $\vartheta_{\rm H}=30^{\circ}$, $35^{\circ}$ and $40^{\circ}$
and distances $R=200$~m and 900~m.  Note that the peak emission is
centered around $\phi_{\rm H} = 0$, as expected from
theory~\cite{HAAAJS1}; variations of $\phi_{\rm H}$ of around $\pm
10^{\circ}$ away from zero do not affect the intensity much, but
thereafter, the intensity falls away rapidly.

Taken in conjunction with Fig.~\ref{expfiga}, the data in
Fig.~\ref{phidata} show that the emission from the superluminal source
is tightly beamed in both the azimuthal and polar directions.  The
observation of a peaked emitted power distribution (see
Fig.~\ref{fcomp}) at two frequencies shows that this effect is
frequency independent, depending instead on the speed of the source
and the geometry of the array.  Once again we contrast this with the
case of a conventional antenna~\cite{vivaldi}; the beaming occurs even
though the projection of the length of the array onto a plane normal
to the observation direction is comparable to or smaller than the
radiation wavelength {\it and} the observation is made at many
hundreds of Fresnel distances.

\begin{figure}[tbp]
   \centering
\includegraphics[height=9cm]{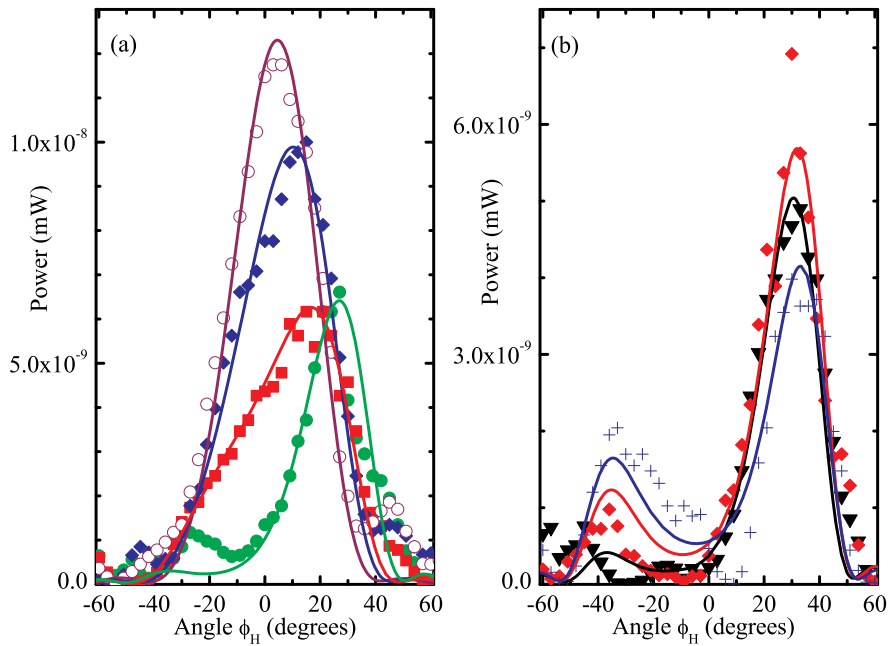}
\caption{(a)~The $\phi_{\rm H}$ dependence of the detected power
(linear scale) for $\vartheta_{\rm H}=40^{\circ}$ (hollow circles),
$35^{\circ}$ (filled diamonds), $30^{\circ}$ (filled squares) and
$25^{\circ}$ (filled circles). Points are data; the curves are the
model (Eq.~\ref{power}) predictions.  (b)~The $\phi_{\rm H}$
dependence of the detected power for $\vartheta_{\rm H}=20^{\circ}$
(inverted triangles), $15^{\circ}$ (diamonds) and $10^{\circ}$
(crosses).  Points are data; the curves are the model predictions
(Eq.~\ref{power}).  In both (a) and (b) $R=400$~m, the frequency is
$f_{+} = (\eta + \Omega)/2 \pi =598.696$~MHz, $v=1.06 c$, $h_{\rm A} =
h_{\rm D} = 5.0$~m.  }
\label{phibeamshape}
\end{figure}

Finally, we remark that the evolution of the beam profile as
$\vartheta_{\rm H}$ and $\phi_{\rm H}$ vary is in good agreement with
the expectations of theory~\cite{HAAAJS1}.  Fig.~\ref{phibeamshape}
shows the the $\phi_{\rm H}$ dependence of the detected power (linear
scale) for several values of $\vartheta_{\rm H}$.  The evolution of
the beamshape, including the development of a precursor (a side
maximum), is well reproduced by the model (Appendix, Eq.~\ref{power}),
using the experimental value of $v/c$.

\section{Nonspherically decaying component of the radiation}
\label{nonspherical}
\subsection{General properties}
The previous section has shown that the angular distribution of the
emission from the experimental source is in good quantitative
agreement with theoretical predictions for a rotating, oscillating
superluminal source (see Appendix)~\cite{HAAAJS1,HAAAJS2}.  However,
the most controversial aspect of the model of
Refs.~\cite{HA3,HAAAJS1,HAAAJS2} has been the prediction of a
component of the emission that decays {\it nonspherically}, {\it
i.e.}\ as $1/R$, rather than $1/R^2$~\cite{HA3A,HA3B}.  
This component is due to
an effect, only possible in the case of an accelerated superluminal
source, whereby waves that were emitted over a finite interval of
emission time arrive at the detector during a much shorter interval of
reception time.  The constructive interference of such superposed
waves is strongest if their emission time interval is centered at the
instant at which their source approaches the detector (along the
radiation direction) with the wave speed (the speed of light) and zero
acceleration.  The locus of the series of points for which this occurs
for a single volume element of the source constitutes the propagating
space curve which we have here called the {\it
cusp}~\cite{HA3,HAAAJS2}.

In this section, we describe the location and the characteristic
features of this emission for a source speed of $v=1.064c$ and the
frequency $f_{+}=(\Omega+\eta)/2\pi=598.696$ MHz; the theoretical
predictions for this speed lead one to expect that the cusp loci will
be found close to $\vartheta_{\rm H} = 20^{\circ}$ at small values of
$\phi_{\rm H}$: For a source that is centered at $\varphi=0$ and an
observation point that lies in the far field, Eq.~49 of
Ref.~\cite{HAAAJS2} reduces to $\varphi_{\rm P}\simeq -3\pi/2$,
$\theta_{\rm P}\simeq\arcsin[c/(r\omega)]$, in which $r\omega\equiv
v$.

\begin{figure}[tbp]
   \centering
\includegraphics[height=9cm]{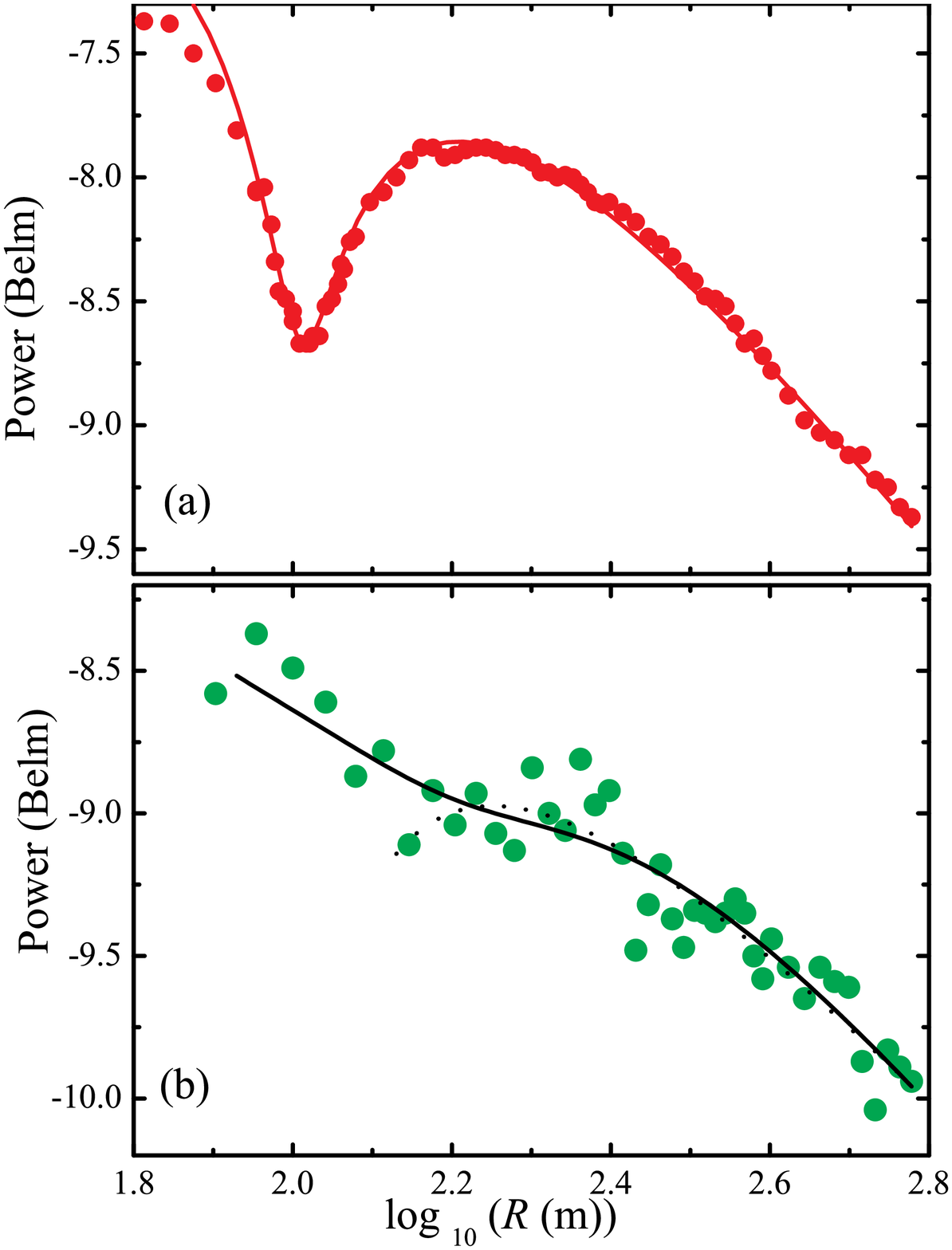}
\includegraphics[height=9cm]{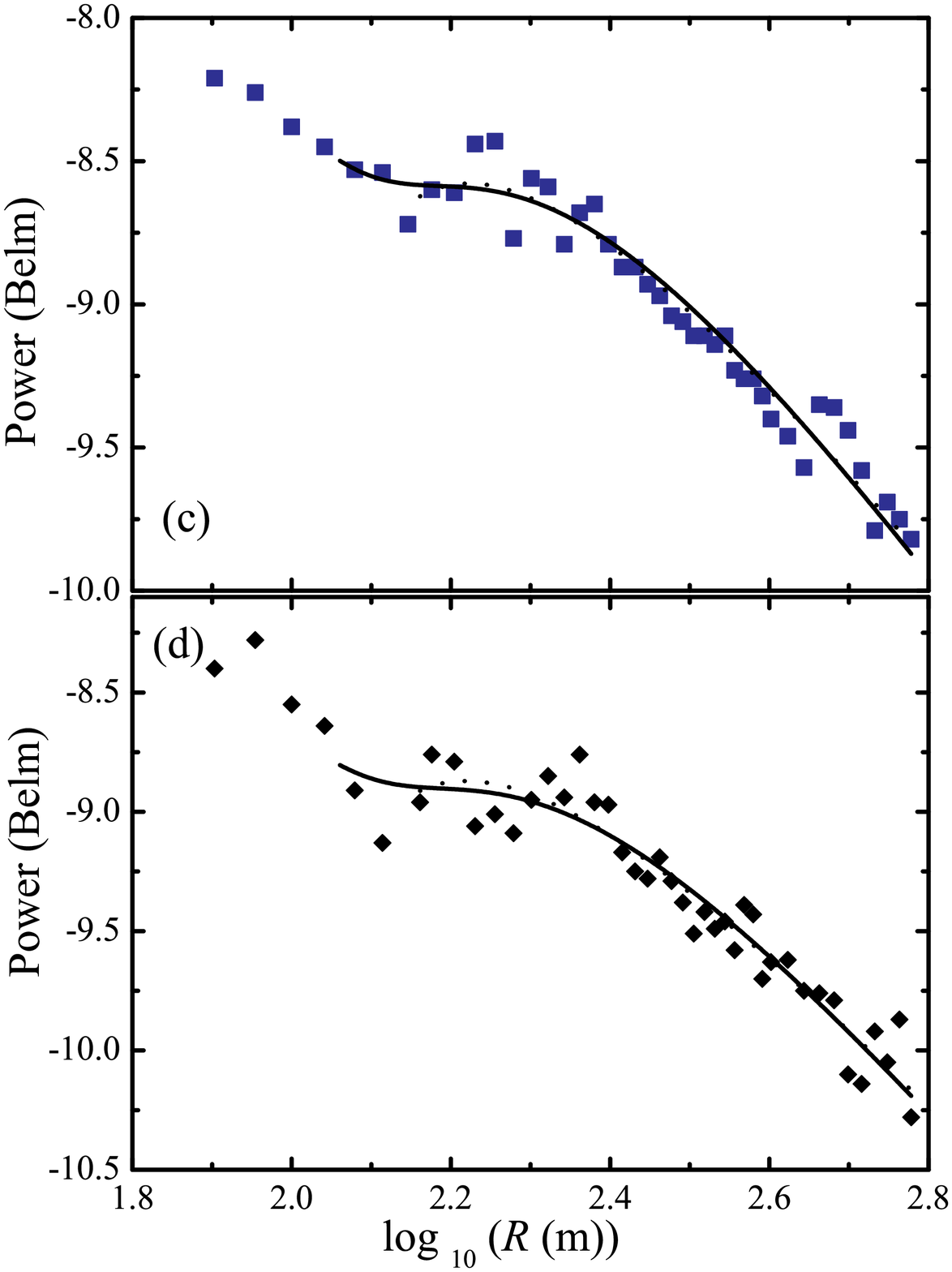}
\caption{(a)~Detected power (logarithmic scale) versus $\log_{10}(R)$,
where $R$ is the array-to-detector distance for a subluminal source,
velocity $v/c=0.875$; $h_{\rm A}=h_{\rm D}= 5.0$~m, $\phi_{\rm H}=0$,
$\vartheta_{\rm H}=35^{\circ}$, $f_{+} = (\eta + \Omega)/2 \pi
=598.696$~MHz.  The points are data; the curve is the interference
model described in the Appendix; the refractive index is $N=1.074$.
(b)~Detected power versus $\log_{10}(R)$ for the superluminal source
close to the expected cusp direction.  The source velocity is
$v/c=1.064$, $h_{\rm A}=h_{\rm D}= 5.0$~m, $\phi_{\rm H}=-5^{\circ}$,
$\vartheta_{\rm H}=20^{\circ}$, $f_{+} = (\eta + \Omega)/2 \pi
=598.696$~MHz.  Data are points; the dotted curve is a fit to
Eq.~\ref{ns} (Appendix).  This model is valid for larger $R$ and
assumes both no beaming and an intensity proportional to
$1/R^{2\kappa}$.  Fit parameters are $2\kappa=0.90$ and $N = 1.074$.
The solid curve represents a fit to Eqs.~\ref{nsq} and \ref{nss} that
uses a Gaussian beam of width $2w$; fit parameters are $2\kappa=1.05$,
$2w=4.18^{\circ}$ and $K=16.6$.  (c)~Detected power versus
$\log_{10}(R)$ to one side of the cusp direction; all experimental
details are the same as in (b) except $\phi_{\rm H}=0$.  Data are
points, the dotted line is a fit to Eq.~\ref{ns} ($2\kappa =1.41$,
$N=1.074$).  The solid curve is a fit to Eqs~\ref{nsq} and \ref{nss}
($2\kappa=1.47$, $2w=6.69^{\circ}$, $K=17.3$).  (d)~Same as (c) but
with $\phi_{\rm H}=-10^{\circ}$. Data are points; the fit to
Eq.~\ref{ns} is dotted ($2\kappa =1.47$, $N=1.074$).  the solid curve
is a fit to Eqs~\ref{nsq} and \ref{nss} ($2\kappa=1.48$,
$2w=6.56^{\circ}$, $K=14.6$). (The data shown here are limited to
$R\leq 600$ m because the runway at the Turweston Aerodrome dips
sharply in its last $300$~m.)  }
\label{fig10}
\end{figure}

As a reference point, Fig.~\ref{fig10}(a) shows the detected power
versus array-to-detector distance $R$, both plotted on logarithmic
scales, for the case of the source moving subluminally; the velocity
is $v/c=0.875$.  Note that the data (points) are very similar to those
shown in Fig.~\ref{intscheme} for the emission of a dipole aerial;
indeed, the data are well fitted using the interference model for a
point-like source and spherical decay described in the Appendix
(curve), especially at larger distances where the finite array size is
much smaller than $R$. This suggests that the emission of the array
running subluminally is entirely ``conventional''; the detected power
obeys the inverse square law (once the effects of interference are
taken into account), and at large distances the source appears
point-like.

Figure~\ref{fig10}(b) shows detected power versus $\log_{10}(R)$ for
the source running superluminally ($v/c=1.064$), for array-to-detector
distances $R$ measured along the expected direction of the cusp loci.
The contrast with Fig.~\ref{fig10}(a) is immediately visible, and
involves three distinct differences:
\begin{enumerate}
\item
The interference fringe visible at around $\log_{10}(R)=2$ in (a) is
absent in (b).
\item
At larger distances, the detected power falls off with distance more
slowly in (b) than in (a).
\item
The data in (b) possess a much greater ``scatter'' than those in
(a). This scatter occurs in spite of the fact that the detected power
is $\sim 50$~dB above the noise floor; moreover, the scatter is to a
large extent reproducible.
\end{enumerate}
The last point can be understood by noting that the cusps represent
emission by those volume elements of the source that travel towards
the detector at zero acceleration and the speed of light, volume
elements that collectively function as a {\it coherent}
source~\cite{HAAAJS2}. The power emitted by a coherent source (such as
a laser) normally contains reproducible variations known as
``speckle''~\cite{speckle}; the reproducible scatter seen in
Fig.~\ref{fig10}(b) is due to just such a phenomenon.

Turning to the other differences between Fig.~\ref{fig10}(a) and (b),
the absence of an interference fringe is expected if the radiation
along this direction is tightly beamed (the loci of the cusps are
expected to have narrow angular width~\cite{HAAAJS1,HAAAJS2}), so that
it is not reflected from the ground close to the array; a detailed
treatment of this effect is given in the Appendix (Eqs.~\ref{nsq} and
\ref{nss}).  The slower decrease in power with increasing distance is
another consequence of the properties of the cusps; we shall return to
the exact exponent describing the decrease below.

\begin{figure}[tbp]
   \centering
\includegraphics[height=9cm]{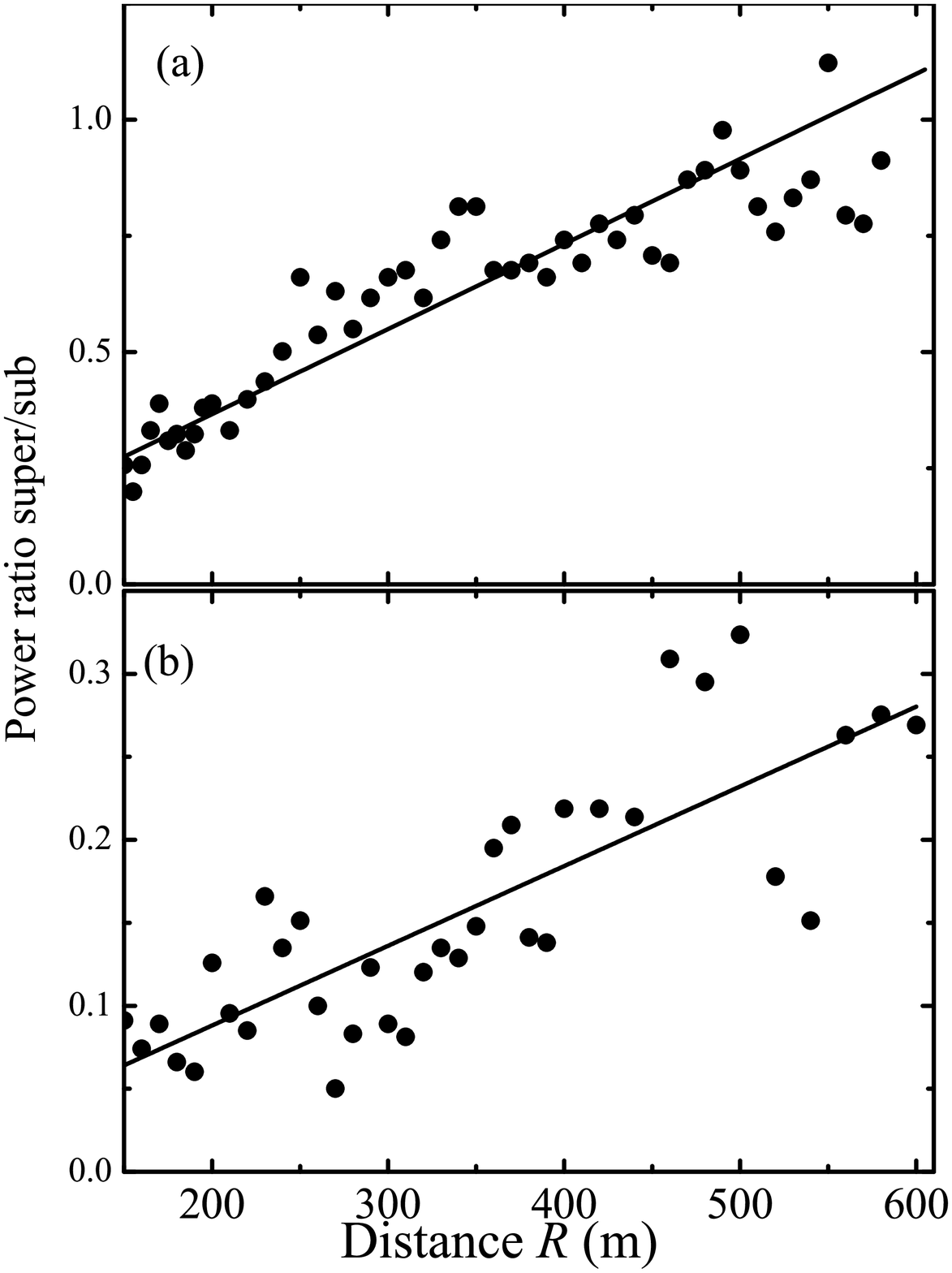}
\includegraphics[height=9cm]{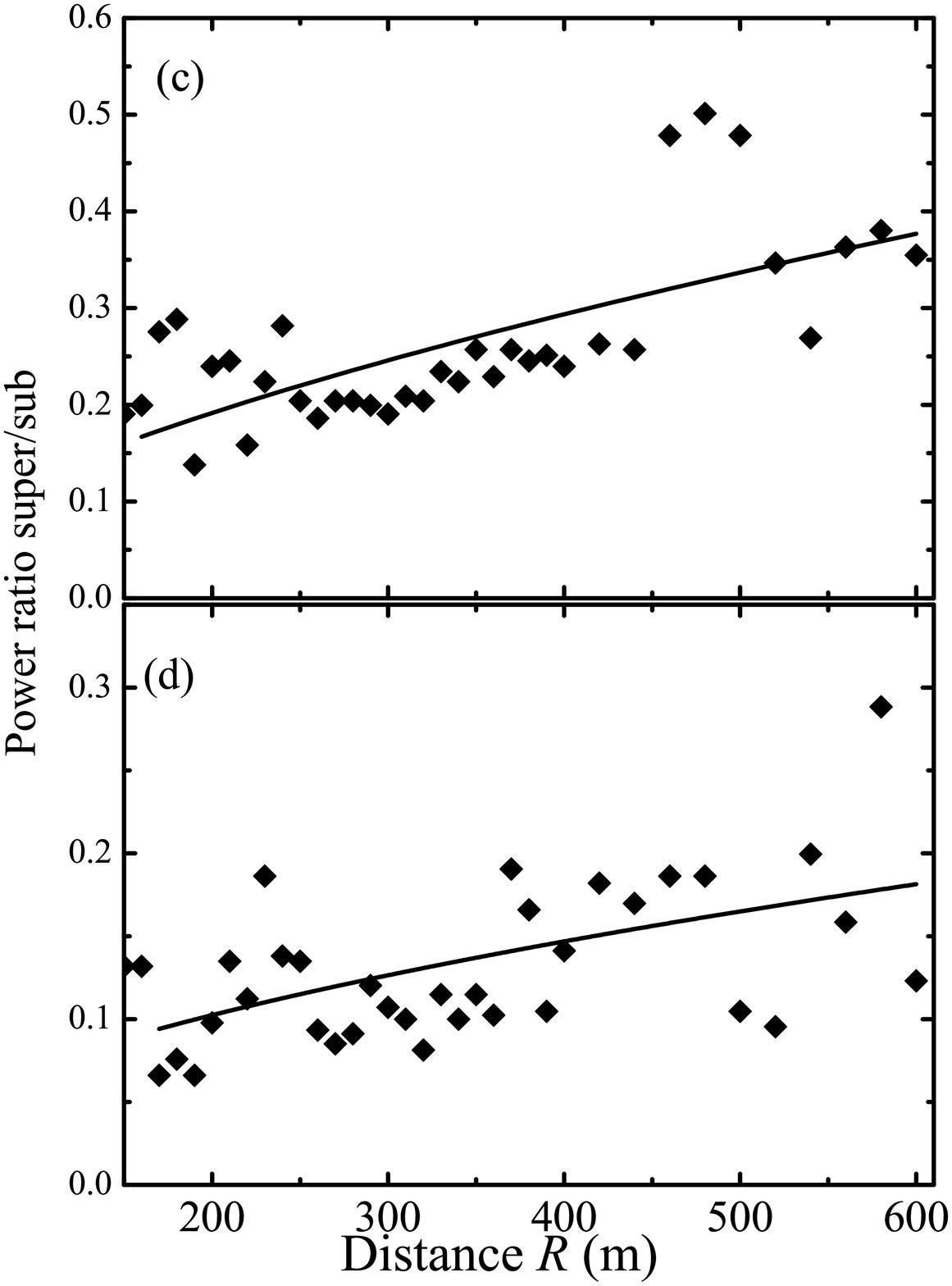}
\caption{(a) and (b): Ratio ${\cal P}^{\rm cusp}/{\cal P}^{\rm sub}$
(see Eq.~\ref{powerrat}) of the detected power with the source running
superluminally (at $v/c=1.064$) to that with the source running
subluminally ($v/c=0.875$) versus $R$, measured on two separate
occasions.  Both data sets follow paths close to the expected
direction of the cusps; $h_{\rm A}=h_{\rm D}= 5.0$~m, $\phi_{\rm
H}=-5^{\circ}$, $\vartheta_{\rm H}=20^{\circ}$, $f_{+} = (\eta +
\Omega)/2 \pi =598.696$~MHz.  Data are points; the lines are fits to
Eq.~\ref{plaw} with $\mu \approx 1$.  A linear fit through the origin
implies that the power in the superluminal case would decline in free
space as $1/R$ (see Eq.~\ref{powerrat}).  (c) and (d) show similar
plots just off the cusp direction; $\varphi_{\rm H} = 0$ in (c), and
$\varphi_{\rm H} =-10^{\circ}$ in (d). All other parameters are the
same as in (a) and (b).  The fits (solid curves) are to Eq.~\ref{plaw}
with $\mu =0.62$ in (c) and $\mu = 0.52$ in (d); {\it i.e.}\ the power
ratio increases more slowly with distance, and cannot be fitted with a
line through the origin.  }
\label{fig11}
\end{figure}

As mentioned above, the cusps occupy only a small angular range;
morover, they are expected to become narrower, the further one goes
from the source. Some idea of this effect is gained by rotating the
array by $5^{\circ}$; whereas Fig.~\ref{fig10}(b) has $\phi_{\rm
H}=-5^{\circ}$, Figs.~\ref{fig10}(c) and (d) show similar data for
$\phi_{\rm H}=0$ and $\phi_{\rm H}=-10^{\circ}$ respectively.  Note
that the small shift in direction has resulted in data that exhibit a
smaller degree of ``scatter'' compared to those in Fig.~\ref{fig10}(b)
(especially at larger $R$), and that the detected power falls off more
rapidly with increasing distance in (c) and (d) than in (b).  The
first observation is entirely expected; once one is off the cusp loci,
the source no longer appears as coherent~\cite{HAAAJS2}, and so the
``speckle'' effects should decrease.  Along directions that
appreciably differ from that of the cusp loci, the emitted power has
the same dependence on distance as in Fig.~\ref{fig10}(a), {\it i.e.}\
as in the case of a subluminal source or a conventional dipole.

Having identified the distinct features of the non-spherically-decaying
component of the radiation, we describe in the following sections 
three experimental methods for determining the exponent $2\kappa$ that
describes the rate at which the intensity of this component decays with
distance $R$ ({\it i.e.}\ intensity~$\propto 1/R^{2\kappa}$).
In each case, we shall find that $2\kappa \approx 1$, in agreement with
theoretical expectations~\cite{HA3,HAAAJS2}.
\subsection{Determination of the rate at which power falls off with
  distance using models of the interference} Comparison of
Figs.~\ref{fig10}(a)-(d) shows that the detected power decreases most
slowly with increasing distance in the case of superluminal velocities
and measurements along the cusp direction (Fig.~\ref{fig10}(b)).
However, obtaining the exponent $2\kappa$ for this decrease is
complicated by the effects of interference from the ground; we have
already seen that even the emission of a simple dipole does not
decrease as $1/R^2$ (Fig.~\ref{intscheme}(b)) in proximity to the
ground.  The part of the radiation corresponding to the cusps provides
a further complication in that it is expected to be tightly
beamed~\cite{HA3,HAAAJS2}; the current experimental machine agrees
with these predictions in that the beam width of of the
non-spherically-decaying component is appreciably (by an order of
magnitude or more) narrower than the beams observed in the
spherically-decaying part of the radiation (as shown in
Figs.~\ref{expfiga}-\ref{phibeamshape}).  A narrow beam will not be
appreciably reflected from the ground when the detector is close to
the array; hence, features due to interference will be suppressed.

Two treatments for this problem are included in the Appendix
(Eqs.~\ref{ns} to \ref{nss}).  If the beam has an angular spread $2w$,
then at distances $R~\gtsim ~h_{\rm A}/w$, a substantial proportion of
the beam will be reflected from the ground; therefore a relatively
simple treatment of interference (Eq.~\ref{ns}) can be used. The only
input parameters are the power law of the decay of the radiation with
distance (intensity~$\propto 1/R^{2\kappa}$) and $N$, the refractive
index of the ground (already constrained to be $N=1.074$ by fitting
the interference fringes of the subluminal emission--see
Fig.~\ref{fig10}(a)).  This method (Fig.~\ref{fig10}(b), dotted curve)
yields a value $2\kappa = 0.90$, {\it i.e.}, an intensity proportional
to $1/R^{0.90}$.

For smaller values of $R$, a more sophisticated fitting method
(Appendix; Eqs.~\ref{nsq}, \ref{nss}) takes account of the beaming by
using a Gaussian beam of angular width $2w$, and also allowing for the
presence of residual spherically-decaying waves close to the array.
This function is shown as a solid curve in Fig.~\ref{fig10}(b); the
fit parameters are $2\kappa=1.05$, $2w=4.18^{\circ}$ and $K=16.6$ (see
Eq.~\ref{nss}).  This implies that the non-spherically-decaying part
of the radiation has an intensity that decays as $1/R^{1.05}$ and that
it is confined to an angular width $\sim 4^{\circ}$. The fact that
$K$, which depends on the geometry and perfection of the source, is
substantially larger than 1 implies that the dominant emission from
the experimental machine is spherically-decaying, with a small
component of non-spherically-decaying emission.

The narrow angular width of the non-spherically-decaying part of the
radiation is emphasised by Figs.~\ref{fig10}(c) and (d), which show
the detected power versus distance for $\phi_{\rm H}=0$ and $\phi_{\rm
H} = -10^{\circ}$ (rather than the $\phi_{\rm H}=-5^{\circ}$ used in
Fig~\ref{fig10}(b)). The power falls off more quickly at large
distances in (c) and (d) than in (b), and this is reflected in the
fits. For example, in (c), the simple model (dotted curve;
Eq.~\ref{ns}) yields $2\kappa =1.41$ (with $N$ constrained to be
$1.074$ to match the fit in Fig.~\ref{fig10}(a)), whereas the more
sophisticated Eqs.~\ref{nsq} and \ref{nss} (solid curve) give
$2\kappa=1.47$, $2w=6.69^{\circ}$ and $K=17.3$.  Both fits suggest an
intensity that is approximately $\propto 1/R^{1.5}$ and the latter is
suggestive of less beaming of the radiation.  Similar considerations
apply to the fits to the data in Fig.~\ref{fig10}(d).

\subsection{Determination of the rate at which power falls off with
  distance using intensity ratios} 
An alternative, model independent
way of determining the exponent $2\kappa$ is to divide the power
detected along the cusp by that detected when the source is run
subluminally under similar conditions.  Such a comparison can only be
made beyond the region containing the interference fringes ({\it
i.e.}\ $R> 150$~m), because (as mentioned above) the cusp has a
distinctly ``beamed'' character, whereas the radiation from the
subluminal source is much more isotropic.  This comparison is shown in
Figs.~\ref{fig11}(a) and (b) for data taken on two separate occasions
with differing weather conditions.  Each data set represents the
detected power ${\cal P}^{\rm cusp}$ along the expected cusp direction
with the source running superluminally ($v/c=1.064$), divided by the
detected power ${\cal P}^{\rm sub}$ with the source running
subluminally ($v/c=0.875$).  In both subluminal and superluminal runs,
all other experimental parameters were identical.

The data were fitted to a function
\begin{equation}
\frac{{\cal P}^{\rm cusp}}{{\cal P}^{\rm sub}} ={\cal C} R^{\mu}
\label{plaw}
\end{equation}
where $\mu$ and ${\cal C}$ are fit parameters. The fits yielded $\mu
\approx 1$, {\it i.e.} a straight line through the origin, implying
that the power ratio is roughly proportional to $R$
(Figs.~\ref{fig11}(a), (b)).  This result can be explained as follows.
Once the interference fringes have been passed, ${\cal P}^{\rm sub}$
will obey an approximate relationship
\begin{equation}
{\cal P}^{\rm sub}(R) \approx \frac{{\cal C}^{\rm sub}}{R^{~\nu}},
\end{equation}
where ${\cal C}^{\rm sub}$ is a constant or a very slowly-varying
function of $R$.  There are many precedents for the use of such power
laws in the empirical functions~\cite{ground,egli,naka} for describing
radio transmission ({\it e.g.}\ Eq.~\ref{eglilaw}).  The straight
lines through the origin seen in Figs.~\ref{fig11}(a),(b) will only
occur if the power ${\cal P}^{\rm cusp}$ follows the relationship
\begin{equation}
{\cal P}^{\rm cusp} \approx \frac{{\cal C}^{\rm cusp}}{R^{~\nu-1}},
\end{equation} 
so that
\begin{equation}
\frac{{\cal P}^{\rm cusp}}{{\cal P}^{\rm sub}} \approx
\frac{{\cal C}^{\rm cusp}}{{\cal C}^{\rm sub}} R.
\label{powerrat}
\end{equation}
This again suggests that the power in the non-spherically-decaying
component of the radiation would decay in free space more slowly
with distance $R$
than $1/R^2$.

Once again, the narrowness of the region occupied by the cusps can be
demonstrated by shifting $\phi_{\rm H}$ by only $5^{\circ}$
(Figs.~\ref{fig11}(c),(d)).  The power ratio increases more slowly
with distance, and the data can no longer be fitted by a straight line
through the origin; instead $\mu = 0.62$ in (c) and $0.52$ in (d) (see
Eq.~\ref{plaw}).
\subsection{Summary}
The various methods of characterizing the rate at which the detected
power falls off with $R$ shown in Figs.~\ref{fig10} and ~\ref{fig11}
suggest that the intensity of the non-spherically-decaying component
is roughly proportional to $1/R$, in agreement with theoretical
predictions~\cite{HA3,HAAAJS2}. Moreover, the relatively small angular
range over which this $1/R$ dependence is observed (compare
Figs.~\ref{fig10}(b) and (c)) emphasises the narrowness of the region
occupied by the cusps.

Note that in the present experiments, the Fresnel distance (Rayleigh
range) is of the order of a few meters. The maximum length of the
array that can be projected onto a plane perpendicular to the
propagation direction is $l=1.74$~m and the shortest wavelength used
is $\lambda=0.50$~m. These together yield a maximum Fresnel distance
$R_{\rm F}=l^2/\lambda=6.0$~m.  There is no parameter other than the
Fresnel distance to limit the range of validity of the above results,
and these, as we have seen, have been tested for $R\gg R_{\rm F}$.

\section{Conclusion}
\label{discussion}
In summary, we have constructed and
tested an experimental implementation of an
oscillating, superluminal polarization
current distribution that undergoes centripetal acceleration.
Theoretical treatments~\cite{HAAAJS1,HAAAJS2} lead one
to expect that the radiation emitted from each volume element 
of such a polarization current will comprise a \v Cerenkov-like
envelope with two sheets that meet along a cusp.
The emission from the experimental machine has been found to be
in good agreement with these expectations, the combined
effect of the volume elements leading to tightly-defined beams
of a well-defined geometry, determined by the source
speed and trajectory. In addition, over a restricted range of angles,
we detect the presence of cusps in the emitted radiation.
These are due to the detection over a short time period
(in the laboratory frame) of radiation emitted over 
a considerably longer period of source time.
Consequently, the
intensity of the radiation at these angles
was observed to decline more
slowly with increasing distance from the source than would the
emission from a conventional antenna.  
The angular distribution of the emitted
radiation and the properties associated with the cusp are 
in good {\it quantitative} agreement with theoretical
models of superluminal sources
once the effect of
reflections from the earth's surface are taken into account.
In particular, the prediction that the beaming and the slow
decay should extend into the far-zone has been tested to
several hundred Fresnel distances.

The above results have implications not only for the interpretation
of observational data on pulsars, which originally motivated the
theoretical investigation of the present emission
process~\cite{HA3,pulsar}, but also for a diverse set of disciplines
and technologies in which these findings have potential practical
applications~\cite{patent}.  One such technology is long-range and
high-bandwidth telecommunications.
    
Since all of the processes involved in generating the polarization
current distribution are linear, and the direction of the emitted
``beam'' only depends on the source speed, it should be simple to use
an array to broadcast several beams in different directions
simultaneously.  This could be achieved by applying voltages which are
sums of terms like Eq.~\ref{volt1}, each with a different speed, to
the electrodes of an array. The beaming is frequency independent, and
so an array could potentially be used to transmit multiple directed
beams at several frequencies. (The very small side lobes, especially in
free space, are a further characteristic of the beaming produced by the
present emission mechanism that facilitates this type of
transmission.)

Of particular practical significance are the properties of the
non-spherically-decaying component of the radiation that we have
observed: the fact that its intensity decays like $1/R$, rather than
$1/R^2$, with the distance $R$ from its source, and that its
beam-width is less than $\sim 5^{\circ}$ at $\sim500$ m.  
This radiation mechanism could thus be used, in
principle, to convey electronic data over very long distances with
very much lower attenuation than is possible with existing sources.
In the case of ground-to-satellite communications, for example, the
above properties imply that either far fewer satellites would be
required for the same bandwidth or each satellite could handle a much
wider range of signals for the same power output.

The excellent correspondence between the experimental data and
theoretical predictions supports the proposal that polarization
current distributions act as superluminal sources of radiation, and
that they can be used for studying and utilizing the novel effects
associated with superluminal electrodynamics. The apparatus described
in this paper represents just one possible geometry and excitation
scheme. In principle, a whole class of superluminal sources are
possible, corresponding, for example, to a rectilinearly accelerated
polarization current~\cite{HA3} (rather than a centripetally
accelerated one), or to a medium that is polarized by impinging laser
beams (rather than by applied voltages), and so on.  It is hoped that
the publication of this paper will stimulate further work along these
lines.

\section{Acknowledgements}
The experimental research was supported by the Engineering and
Physical Sciences Research Council (UK), under grant GR/M52205/01.
The parts of this work carried out at NHMFL Los Alamos were performed
under the auspices of the National Science Foundation, Department of
Energy and the State of Florida. Gregory S. Boebinger, Roger Cowley,
Mike Glazer, Mehdi Golshani, Alex H. Lacerda, Donald
Lynden-Bell, Chuck Mielke, John Rodenburg and Colin Webb 
are thanked for their great
encouragement in this undertaking.  We should like to express our
gratitude to James Analytis, Alimamy Bangura,
Amalia Coldea, Paul Goddard, Gavin
Morley, Casper Muncaster and Alessandro Narduzzo for
their physical assistance with the experiments and to Steve Blundell,
Scott Crooker, Albert Migliori and 
Dwight Rickel for very fruitful discussions.  In
addition, we are grateful to Dr David Owen (Turweston Aerodrome), Mr
Frederick Jenkinson, Mr Walter Sawyer (Keeper of the Oxford University
Parks) and the Trustees of the Wyre-Lune Sanctuary for their
generosity in allowing us access to the land used for range tests.
Finally, we thank Heather Bower, Ross McDonald and James and Dorothy
Singleton for assisting in the transportation of the apparatus and
people involved in the experiments. AA is supported by the Royal Society.

\section{Appendix: procedures used in modeling and fitting the experimental data}
\subsection*{Introduction}
This Appendix contains the procedures used to model and fit the
experimental data.  It is necessary to be rather rigorous in treating
the radiation from a superluminal source and the effects of
interference from the ground, and so we have described the
mathematical techniques in some detail, a level of detail that would
impede the discussion of the experimental data if it were included in
the main body of the paper.  Moreover, some effects described in the
Appendix, such as interference caused by reflections from the ground,
feature in the interpretation in all sections of the paper. It is
therefore logical to collect all of the discussions of these phenomena
in one place.

The Appendix is divided into two main sections, treatment of
interference due to reflections from the ground (see
Fig.~\ref{intscheme}), and use of the theory contained in
Refs.~\cite{HAAAJS1,HAAAJS2} to model the emissions of the
experimental array (and in particular the beaming of the radiation).
Finally a shorter section deals with the robustness of the (small
number of) fitting parameters used to model the data.

\subsection*{Interference caused by reflections from the ground} 
It is well known from the telecommunications
literature~\cite{ground,egli,naka,Stratton} that interference is a
significant complication in the propagation of radio waves close to
the ground; the detected signal is the result of the interference
between the direct and the reflected waves, {\em i.e.}\ between the
waves that propagate directly from the source to the detector and the
waves that reach the director after reflection from the ground.  In
the current experiments, this effect has been characterized by
observing the emission from both a small dipole aerial and the array
of the source driven at a {\it subluminal} velocity.

\vspace{3mm}
\noindent
{\bf Simple treatment of the effects of interference on the radiation of a dipole aerial}

\vspace{3mm}
\noindent
We treat the dipole aerial as a point source at $x=0,~y=0,~z=h_{\rm
A}$ emitting radiation of wavenumber $k$ in the presence of a medium,
index of refraction $N$, occupying $z\leq 0$. The detector is assumed
to be at the height $h_{\rm D}$ and a distance $R$ away from the
source (see Fig.\ 4(a)).
 
Had it not been deflected by the ground, the reflected ray that passes
through the observation point P would have passed through the mirror
image P${}^\prime$ of P with respect to the ground.  Thus the angle
$\alpha$ that the reflected ray makes with the normal to the ground
${\hat{\bf e}}_{\rm g}$, and the distance $R_{\rm r}$ that the
reflected wave traverses to reach the detector are respectively given
by
\begin{equation}
\alpha=\arctan\{[R^2-(h_{\rm D}-h_{\rm A})^2]^{1\over2}/(h_{\rm D}+h_{\rm A})\},  
\end{equation}
and 
\begin{equation}
R_{\rm r}=(R^2+4h_{\rm A}h_{\rm D})^{1\over2}
\end{equation}
(see Fig.\ 4(a)).  When $R$ is large enough for the curvature of the
wave fronts to be negligible, the reflected and the incident waves are
related via the Fresnel coefficients:
\begin{equation}
{\bf E}_{\rm reflected}=q_\perp{\bf E}_{\rm incident}, 
\label{refla}
\end{equation}
or
\begin{equation}
{\bf B}_{\rm reflected}=q_\parallel{\bf B}_{\rm incident}
\label{reflb}
\end{equation}
(see~\cite{Stratton}).  The vectors ${\bf E}$ and ${\bf B}$ in these
relations are the electric and magnetic fields of the incident and
reflected waves and
\begin{equation}
q_\perp={{\cos\alpha-(N^2-\sin^2\alpha)^{1\over2}}\over{\cos\alpha+(N^2-\sin^2\alpha)^{1\over2}}},
\label{qperp}
\end{equation}
\begin{equation}
q_\parallel={{N^2\cos\alpha-(N^2-\sin^2\alpha)^{1\over2}}\over{N^2\cos\alpha+(N^2-\sin^2\alpha)^{1\over2}}}
\label{qparallel}
\end{equation}
are the Fresnel coefficients for ${\bf E}_{\rm incident}$
perpendicular or ${\bf E}_{\rm incident}$ parallel to the plane of
incidence, respectively; the magnetic permeability of the ground is
here set equal to unity.  (Note that the index of refraction $N$ and
so the coefficients $q_\perp$ and $q_\parallel$ are in general
complex.  However, for waves of frequency $f$, the imaginary parts of
these quantities are negligible when $2\sigma/(\epsilon f)\ll 1$,
where $\sigma$ and $\epsilon$ are the conductivity and the dielectric
constant of the reflecting medium, respectively~\cite{Stratton}.  The
latter condition appears to encompass all of the experiments described
in this paper.)

Hence, in the case of the spherically-decaying radiation that is
generated by a point-like aerial in the vicinity of the ground, the
observed power at the detector is given, to within a constant of
proportionality, by
\begin{equation}
{\cal P}^{\rm s}_{\parallel,\perp} \propto 
\Big\vert {\exp({\rm i}kR)\over R}+
q_{\parallel,\perp}{\exp({\rm i}kR_{\rm r})\over R_{\rm r}}\Big\vert^2,
\label{intsummary}
\end{equation}
where the subscripts in ${\cal P}^{\rm s}_\parallel$ and ${\cal
P}^{\rm s}_\perp$ refer to the common polarizations of the incident
and reflected waves: parallel or perpendicular to the plane of
incidence.  By measuring ${\cal P}^{\rm s}_\parallel$ or ${\cal
P}^{\rm s}_\perp$ as a function of the distance $R$ and fitting the
resulting data to the above equation, it is possible to determine $N$
for the ground (i.e.\ the airfield runway) under a particular set of
weather conditions (see Fig.~4(b)).  This is then used in
Eqs.~\ref{ns} and \ref{nss} (see below) to model the emission of the
superluminal source.

\vspace{3mm}
\noindent
{\bf The effects of interference on the non-spherically-decaying 
component of the radiation from a superluminal source}

\vspace{3mm}
\noindent
The radiation by a superluminal source consists of two components: a
spherically-decaying component whose intensity has the dependence
$R^{-2}$ on $R$ (as does any other radiation) and a
nonspherically-decaying component.  To treat the latter component in
the presence of reflection from the ground, we rewrite
Eq.~\ref{intsummary} as follows:
\begin{equation}
{\cal P}^{\rm ns}_{\parallel,\perp} \propto 
\Big\vert {\exp({\rm i}kR)\over R^\kappa}+
q_{\parallel,\perp}{\exp({\rm i}kR_{\rm r})\over {R_{\rm r}}^\kappa}\Big\vert^2.
\label{ns}
\end{equation}
In other words, the exponent $\kappa$, describing the rate at which
the amplitude of each component (direct and reflected) diminishes with
increasing $R$, has become an adjustable parameter.  The theoretically
predicted value of $\kappa$ for a centripetally accelerated
superluminal source is $\textstyle{1\over2}$ (see~\cite{HA3,HAAAJS2}).

\begin{figure}[tbp]
   \centering
\includegraphics[height=4cm]{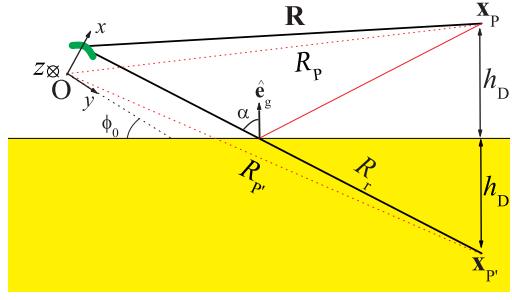}
\caption{Relationship between the positions of the source, the
detector and the notional ``image'' detector.}
\label{fig12}
\end{figure}

Eq.~\ref{ns} will only be valid if the amplitudes of the radiation
traveling via the direct and reflected paths are similar.  However, in
the present experiment, the beam-width of the nonspherically-decaying
component is appreciably (by a factor of 10 or more) smaller than that
of the spherically-decaying component.  This results in a substantial
difference between the amplitudes of the direct and reflected rays
when the detector is close to the source.  (It is this phenomenon that
is responsible for suppressing the interference fringes in
Figs.~\ref{fig10}(b), (c) and (d).)  Unless $R$ is large, therefore,
the Fresnel coefficients in the above expression should be modified to
take account of beaming.  When modeling the data for ranges $R~\gtsim
~150$~m, we have consequently replaced the coefficients
$q_{\parallel,\perp}$ by
\begin{equation}
Q_{\parallel,\perp}= q_{\parallel,\perp}\exp[-(\textstyle{1\over2}\pi-\alpha)^2/w^2],
\label{nsq}
\end{equation}
where $\alpha$ is the angle between an incident ray and the normal to
the reflecting surface (see Fig.~\ref{fig12}) and $w$ represents the
Gaussian half-width of the beam.  The exponential factor multiplying
the Fresnel coefficients greatly reduces the amplitudes of the
reflected waves for which $\alpha$ differs from $\pi/2$ by more than
$w$.

As $R$ tends to a value much larger than $c/\omega$ along this narrow
beam, the nonspherically-decaying component of the radiation dominates
over the conventional component.  At shorter distances from the
source, however, the detected power receives contributions from both
of these components. If the fitting includes data for relatively small
values of $R$ (see {\it e.g.} the solid curves in Figs.~\ref{fig10}(b)
and (c)), both spherically- and non-spherically-decaying contributions
must be combined as follows:
\begin{equation}
{\cal P}_{\parallel,\perp}\propto 
\Big\vert {\exp[{\rm i}(kR+\pi/4)]\over R^\kappa}+K{\exp({\rm i}kR)\over R}
+Q_{\parallel,\perp}\Big[{\exp[{\rm i}(kR_{\rm r}+\pi/4)]\over R_{\rm r}^\kappa}
+K{\exp[{\rm i}
(kR_{\rm r})]\over R_{\rm r}}\Big]\Big\vert^2.
\label{nss}
\end{equation}
Here, the constant $K$ stands for the relative strengths of the
sources of the two components, and $\pi/4$ is the phase difference
betweem them (see~\cite{HAAAJS1,HAAAJS2}).

\vspace{3mm} 
\noindent
{\bf Extension to a volume source:  description of the direct and reflected rays 
in terms of the experimentally measured coordinates}

\vspace{3mm}
\noindent
The modeling of the data is based on the same coordinates as those
used in the theory papers~\cite{HAAAJS1,HAAAJS2}, shown in
Fig.~\ref{fig3}(c,d).  In these coordinates, the source consists of
the arc $-\pi/36<\varphi<\pi/36$ of a circular strip, centered on the
origin, with the radius $a=10.025$~m and the thickness $\Delta
r=0.05$~m. The source velocity points in the direction of increasing
$\varphi$ (see Fig.~\ref{fig3}(d); its angular velocity points in the
direction of positive $z$).  Since the position of the source in
Fig.~\ref{fig3}(a) is such that its velocity points from the left to
the right, the spherical coordinates $(R_P,\theta_P,\varphi_P)$ of
Refs.~\cite{HAAAJS1,HAAAJS2} have to be based, in the present case, on
a Cartesian system whose $z$-axis points downward, as in
Fig.~\ref{fig3}(c).  (The midpoint of the source arc lies on the
$x$-axis of this underlying Cartesian system; see Fig.~\ref{fig3}(c).)

For comparing the data with the theoretical predictions, we need to
express $(R_{\rm P},\theta_{\rm P},\varphi_{\rm P})$ in terms of the
experimentally measured coordinates $(R,\vartheta_{\rm V},\phi_{\rm
V})$ and $(R,\vartheta_{\rm H},\phi_{\rm H})$ of Figs.~\ref{fig3}(a)
and \ref{fig3}(b).
 
The Cartesian system $(x^\prime,y^\prime,z^\prime)$ corresponding to
the spherical coordinates $(R,\vartheta_V,\phi_V)$ of
Fig.~\ref{fig3}(a) differs from the Cartesian system $(x,y,z)$
corresponding to the spherical coordinates $(R_{\rm P},\theta_{\rm
P},\varphi_{\rm P})$ of Fig.~\ref{fig3}(c) in that its origin is
shifted by the array radius $a$ along the positive $x$-axis, its $x$
and $y$ axes are interchanged, and its $z$-axis is inverted:
\begin{equation}
x^\prime=y,\quad y^\prime=x-a,\quad z^\prime=-z.
\label{eqone}
\end{equation}  
The Cartesian and polar coordinates of the observation point P in the
reference frame shown in Fig.~\ref{fig3}(a) are related by
\begin{equation}
x^\prime=R\cos\vartheta_{\rm V}\cos\phi_{\rm V},\quad y^\prime=
R\cos\vartheta_{\rm V}\sin\phi_{\rm V},
\quad z^\prime=R\sin\vartheta.
\end{equation}
These, together with the corresponding coordinates of P in the
reference frame shown in Figs.~\ref{fig3}(c), (d),
\begin{equation} 
x=R_{\rm P}\sin\theta_{\rm P}\cos\varphi_{\rm P},\quad 
y=R_{\rm P}\sin\theta_{\rm P}\sin\varphi_{\rm P},\quad 
z=R\cos\theta_{\rm P},
\end{equation}
yield the first required transformation:
\[
R_{\rm P}=(R^2+a^2+2aR\cos\vartheta_{\rm V}\sin\phi_{\rm V})^{1/2},
\]
\[
\theta_{\rm P}=\arccos\Big(-{R\over R_{\rm P}}\sin\vartheta_{\rm V}\Big),
\]
\begin{equation}
\varphi_{\rm P}=-3\pi/2-\arctan\Big(\tan\phi_{\rm V}+
{a\over R\cos\vartheta_{\rm V}\cos\phi_{\rm V}}\Big).
\label{transform1a}
\end{equation}

In Fig.~\ref{fig3}(b), the position vector of the observation point P
is projected onto the $(x^\prime,z^\prime)$ plane of the shifted
Cartesian system $(x',y',z')$, so that
\begin{equation}
x^\prime=R\cos\phi_{\rm H}\cos\vartheta_{\rm H},\quad y^\prime=R\sin\phi_{\rm H},
\quad z^\prime=R\cos\phi_{\rm H}\sin\vartheta_{\rm H}.
\label{horiz}
\end{equation}  
Rewriting these expressions for $(x^\prime,y^\prime,z^\prime)$ in
terms of $(R_{\rm P},\theta_{\rm P},\varphi_{\rm P})$, via $(x,y,z)$,
and solving for $R_{\rm P}$, $\theta_{\rm P}$ and $\varphi_{\rm P}$,
we obtain the second required transformation:
\[
R_{\rm P}=(R^2+a^2+2aR\sin\phi_{\rm H})^{1/2},
\]
\[
\theta_{\rm P}=\arccos\Big(-{R\over R_{\rm P}}
\cos\phi_{\rm H}\sin\vartheta_{\rm H}\Big),
\]
\begin{equation}
\varphi_{\rm P}=-3\pi/2-\arctan
\Big({\tan\phi_{\rm H}\over\cos\vartheta_{\rm H}}+
{a\over R\cos\phi_{\rm H}\cos\vartheta_{\rm H}}\Big).
\end{equation}

Having expressed the position vector ${\bf x}_{\rm P}=(R_{\rm
P},\theta_{\rm P},\varphi_{\rm P})$ of the detector in terms of the
experimentally measured coordinates, we now know how to determine the
characteristics of the direct ray ${\bf x}_{\rm P}-{\bf x}$ that links
each source element ${\bf x}=(r,\varphi,z)$ of the array to the
observation point P.  The next step is to determine the corresponding
characteristics of the reflected rays that reach P from various volume
elements of the array.  As we have already seen in the treatment of a
simple dipole aerial, both the angle $\alpha$ that a given reflected
ray makes with the normal to the ground, ${\hat{\bf e}}_g$, and the
distance $R_r$ that this wave traverses to reach the detector at P
(see Fig.~\ref{fig12}) can be easily inferred from the position vector
${\bf x}_{\rm P'}$ of the image P${}^\prime$ of P with respect to the
ground.  We proceed to determine ${\bf x}_{\rm P'}$ in terms both of
the coordinates $(R,\vartheta_V,\phi_V)$ and of the coordinates
$(R,\vartheta_H,\phi_H)$.

Suppose that the plane of the array is vertical, as in
Fig.~\ref{fig3}(a), and that the upward-pointing normal to the ground
at the point of reflection, ${\hat{\bf e}}_{\rm g}$, makes the angle
$\phi_0$ with the positive $x$-axis of the $(x,y,z)$ frame (whose
$z$-axis is in this case parallel to the ground), {\em i.e.}\
\begin{equation}
{\hat{\bf e}}_{\rm g}=\cos\phi_0{\hat{\bf e}}_x-\sin\phi_0{\hat{\bf e}}_y.
\label{normal}
\end{equation}
(For a smooth and perfectly horizontal ground, $\phi_0$ is related to
$\phi_{\rm V}$ via $\phi_0-\phi_{\rm V}=\arcsin[(h_{\rm A}-h_{\rm
D})/R]$.)  Then the position vector, in the $(x,y,z)$ frame, of the
image P${}^\prime$ of P across the earth's surface can be written as
\begin{equation}
{\bf x}_{\rm P'}={\bf R}+a{\hat{\bf e}}_x-2h_{\rm D}{\hat{\bf e}}_{\rm g},
\label{tfmntwo}
\end{equation}
where ${\bf R}$ is the position vector of P in the $(x',y',z')$ frame,
the frame whose origin lies at the midpoint of the array (see
Figs.~\ref{fig3}(c) and \ref{fig12}).

The following expressions for ${\bf x}_{\rm P'}$ and ${\bf R}$ in
terms of their components in the $(x,y,z)$ and the $(x',y',z')$
frames,
\begin{equation}
{\bf x}_{\rm P'}=R_{\rm P'}\sin\theta_{\rm P'}(\cos\varphi_{\rm P'}{\hat{\bf e}}_x
+\sin\varphi_{\rm P'}{\hat{\bf e}}_y)+R_{\rm P'}\cos\theta_{\rm P'}{\hat{\bf e}}_z,
\label{tfmnthree}
\end{equation}
\begin{equation}
{\bf R}=R\cos\vartheta_{\rm V}(\cos\phi_{\rm V}{\hat{\bf e}}_{x'}
+\sin\phi_{\rm V}{\hat{\bf e}}_{y'})+R\sin\vartheta_{\rm V}{\hat{\bf e}}_{z'},
\end{equation}
together with the relationships
\begin{equation}
{\hat{\bf e}}_{x'}={\hat{\bf e}}_y,
\quad {\hat{\bf e}}_{y'}={\hat{\bf e}}_x,
\quad {\hat{\bf e}}_{z'}=-{\hat{\bf e}}_z
\label{bases}
\end{equation}
between the base vectors of these two frames (see Eq.~\ref{eqone})
therefore yield
\[
R_{\rm P'}=[R^2+a^2+(2h_{\rm D})^2-4ah_{\rm D}\cos\phi_0
+2aR\cos\vartheta_{\rm V}\sin\phi_{\rm V}-4h_{\rm D}R\cos\vartheta_{\rm V}\sin(\phi_{\rm V}-\phi_0)]^{1/2},
\]
\[
\theta_{\rm P'}=\arccos\Big(-{R\over R_{\rm P'}}\sin\vartheta_{\rm V}\Big),
\]
\begin{equation}
\varphi_{\rm P'}=-3\pi/2-\arctan\Big({{R\cos\vartheta_{\rm V}\sin\phi_{\rm V}
+a-2h_{\rm D}\cos\phi_0}\over{R\cos\vartheta_{\rm V}\cos\phi_{\rm V}+2h_{\rm D}\sin\phi_0}}\Big),
\label{tfmnfour}
\end{equation}
for the spherical coordinates of the image point P${}^\prime$ in terms
of the angles $(\vartheta_{\rm V},\phi_{\rm V})$ of
Fig.~\ref{fig3}(a).

To determine ${\bf x}_{\rm P'}$ in terms of the coordinates
$(R,\vartheta_H,\phi_H)$ of Fig.~\ref{fig3}(b), suppose that the
upward-pointing normal to the ground, at the point of reflection,
makes the angle $\vartheta_0$ with the negative $z$-axis of the
$(x,y,z)$ frame (whose $x$-axis is in this case parallel to the
ground), i.e.\
\begin{equation}
{\hat{\bf e}}_{\rm g}=-\sin\vartheta_0{\hat{\bf e}}_y-\cos\vartheta_0{\hat{\bf e}}_z.
\label{norman}
\end{equation}
(Here, too, $\vartheta_0$ is related to $\vartheta_{\rm H}$ via
$\vartheta_0-\vartheta_{\rm H}=\arcsin[(h_{\rm A}-h_{\rm D})]/R$ when
the ground is smooth and horizontal.)

Inserting this alternative expression for ${\hat{\bf e}}_{\rm g}$ in
Eq.~\ref{tfmntwo} and using Eqs.~\ref{tfmnthree},~\ref{bases}, and the
following version of Eq.~\ref{horiz},
\begin{equation}
{\bf R}=R\cos\phi_{\rm H}(\cos\vartheta_{\rm H}{\hat{\bf e}}_{x'}
+\sin\vartheta_{\rm H}{\hat{\bf e}}_{z'})+R\sin\phi_{\rm H}{\hat{\bf e}}_{y'},
\end{equation}
we obtain
\[
R_{\rm P'}=[R^2+a^2+(2h_{\rm D})^2+2aR\sin\phi_{\rm H}
-4h_{\rm D}R\cos\phi_{\rm H}\sin(\vartheta_{\rm H}-\vartheta_0)]^{1/2},
\]
\[
\theta_{\rm P'}=\arccos\Big(-{{R\cos\phi_{\rm H}\sin\vartheta_{\rm H}
-2h_{\rm D}\cos\vartheta_0}\over R_{\rm P'}}\Big),
\]
\begin{equation}
\varphi_{\rm P'}=-3\pi/2-\arctan\Big({{R\sin\phi_{\rm H}+a}
\over{R\cos\phi_{\rm H}\cos\vartheta_{\rm H}
+2h_{\rm D}\sin\vartheta_0}}\Big),
\label{tfmnfive}
\end{equation}
for the spherical coordinates of the image point P${}^\prime$, with
respect to the $(x,y,z)$ frame in terms of the angles
$(\vartheta_H,\phi_H)$ of Fig.~\ref{fig3}(b).

\subsection*{Theoretical treatment of the spherically-decaying
component of the emission from the experimental array}
\label{beamingmodel}
We now describe the use of the theory of Refs.~\cite{HAAAJS1,HAAAJS2}
to model the emissions of the experimental array.  Our purpose is to
predict the beaming of the {\it spherically-decaying} component of the
radiation; relevant data are shown in
Figs.~\ref{expfiga}-\ref{phibeamshape}.  (The detailed behavior of the
nonspherically-decaying component will be left for a future work.)
The same mathematical notation as that in Refs.~\cite{HAAAJS1,HAAAJS2}
is used; for consistency with the original papers, we employ Gaussian
electromagnetic units.
 
We commence with the following far-field expressions for the
electromagnetic fields {\bf E} and {\bf B} in the absence of
boundaries [Eqn.~(14) of Ref.~\cite{HAAAJS1}]:
\begin{equation}
{\bf E}\simeq{1\over c^2}\int {\rm d}^3x\,{\rm d}t\,
{\delta(t_{\rm P}-t-\vert{\bf x}_{\rm P}-{\bf x}\vert/c)\over\vert{\bf x}_{\rm P}-{\bf x}\vert}{\hat{\bf n}}{\bf\times}
\Big({\hat{\bf n}}{\bf\times}
{\partial{\bf j}\over\partial t}\Big)~~~{\rm ;}~~~
{\bf B}={\hat{\bf n}}\times{\bf E}.
\label{haeqn14}
\end{equation}
Here $t_{\rm P}$ is the time at which the field is measured at the
observation point P, and ${\bf x}$ and $t$ are the position of a
source element and the retarded time, respectively; ${\hat{\bf
n}}={\bf x}_{\rm P}/\vert{\bf x}_{\rm P}\vert$ is a unit vector along
the direction linking the source that is localized in the vicinity of
the origin and the observer at P.  For the polarization ${\bf P}$
described in Eq.~\ref{haeq7}, whose distribution pattern both rotates
(with an angular frequency $\omega$) and oscillates (with a frequency
$\Omega$), the source term in the above expression is related to the
cylindrical components $(s_r,s_\varphi,s_z)$ of the vector ${\bf s}$
that designates the direction of the current density ${\bf
j}=\partial{\bf P}/\partial t$ by
\begin{eqnarray}
{\hat{\bf n}}{\bf\times}({\hat{\bf n}}{\bf\times}\partial{\bf j}/\partial t)
=\textstyle{1\over4}\omega^2&\sum_{\mu=\mu_\pm}\mu^2\exp[-{\rm
    i}(\mu{\hat\varphi}-\Omega\varphi/\omega)]
\big\{[s_\varphi\cos\theta_P\sin(\varphi-\varphi_P)\cr
&-s_r\cos\theta_P\cos(\varphi-\varphi_P)+s_z\sin\theta_P]{\hat{\bf
    e}}_\perp+
[s_r\sin(\varphi-\varphi_P)\cr
&+s_\varphi\cos(\varphi-\varphi_P)]{\hat{\bf e}}_\parallel\big\}
+\{m\to-m,\Omega\to -\Omega\},\qquad&~~~
\label{haeqn18}
\end{eqnarray}
where the cylindrical coordinates $(r,\varphi,z)$ and 
$(r_{\rm P},\varphi_{\rm P},z_{\rm P})$ of ${\bf x}$ and ${\bf x}_{\rm P}$ 
are with respect to the $(x,y,z)$ frame shown in 
Fig.~\ref{fig3}(c), ${\hat\varphi}\equiv\varphi-\omega t$, 
$\mu_\pm\equiv(\Omega/\omega)\pm m$, ${\hat{\bf e}}_\parallel={\hat{\bf e}}_{\varphi_ P}$ 
(which is parallel to the plane of rotation) and 
${\hat{\bf e}}_\perp={\hat{\bf n}}\times{\hat{\bf e}}_\parallel$ are a pair 
of unit vectors normal to the radiation direction ${\hat{\bf n}}$, and 
the symbol $\{m\to-m,\Omega\to-\Omega\}$ designates a term exactly 
like the one preceding it but in which $m$ and $\Omega$ are everywhere 
replaced by $-m$ and $-\Omega$, respectively.  This is Eq.~(18) of Ref.~\cite{HAAAJS1}.

We are interested in the context of the present measurements in the
radiation that is polarized parallel to the plane of the array; note
also that the component $s_{\varphi}$ is zero in the experimental
array described in the current paper.  Hence, the relevant component
of the source term in Eq.~\ref{haeqn18} is
\begin{equation}
{\hat{\bf e}}_\parallel\cdot[{\hat{\bf n}}\times({\hat{\bf n}}
\times\partial{\bf j}/\partial t)]=\textstyle{1\over2}s_r\sin(\varphi-\varphi_P)\Re\Big\{ 
\sum_{\pm}(\Omega\pm m\omega)^2\exp\{{\rm i}[(\Omega\pm m\omega)t\mp m\varphi]\}\Big\},
\label{eqna}
\end{equation}
where $\Re\{Z\}$ denotes the real part of $Z$.  The fact that the
array has the form of an arc means that the domain of integration in
Eq.~\ref{haeqn14} consists, in the present case, of the cylindrical
strip $-5^{\circ}<\varphi< 5^{\circ}$, $10.00$~m$<r<10.05$~m,
$-0.005$~m $<z<$ $0.005$~m, that bounds the volume occupied by the
array (see Fig.~\ref{fig1}).  Since the $r$ and $z$ dimensions of this
volume are appreciably shorter than the wavelengths at which the
radiation is generated and measured in the present experiments
($0.50$~m and $0.59$~m), the integration with respect to these two
coordinates may be omitted when evaluating the integral in
Eq.~\ref{haeqn14} without introducing any significant error.  (It must
be emphasized that this approximation may {\it only} be used in the
case of the spherically-decaying component of the emission; it is
necessary to include the full volume of the source when calculating
the nonspherically-decaying component.  For the current
purposes--modeling the spherically-decaying data of
Figs.~\ref{expfiga}-\ref{phibeamshape}--the approximation is
adequate.)

Inserting Eq.~\ref{eqna} in Eq.~\ref{haeqn14} and performing the
trivial integration with respect to $t$, we find that the electric
field of the radiation that is detected at either the frequency
$f_+=\Omega+m\omega\equiv n_+\omega$ or the frequency
$f_-=\vert\Omega-m\omega\vert\equiv \vert n_-\vert\omega$ (see
Table~\ref{table1}~\cite{HAAAJS1,HAAAJS2}) is given by
\begin{equation}
{\bf E}(f_\pm)\propto\Re\Big\{\exp(-{\rm i}n_\pm\omega t_{\rm P})      
\int_{-\pi/36}^{\pi/36} d\varphi\, \sin(\varphi-\varphi_{\rm P})
\exp[{\rm i}(n_\pm{\hat R}\pm m\varphi)]/{\hat R}\Big\}{\hat{\bf e}}_{\varphi_{\rm P}}
\label{eqnb}
\end{equation}
to within a constant of proportionality depending on the source strength.  Here,
\begin{equation} 
{\hat{R}}\equiv\vert{\bf x}_{\rm P}-{\bf x}\vert\omega/c=[{{\hat R}_{\rm P}}^2+{\hat r}^2-2{\hat r}{\hat R}_{\rm P}\sin\theta_P\cos(\varphi-\varphi_P)]^{1\over2},
\end{equation}
in which ${\hat r}\equiv r\omega/c$ is the source speed in units of 
$c$, ${\hat R}_{\rm P}\equiv R_{\rm P}\omega/c=({r_{\rm P}}^2+{z_{\rm P}}^2)^{1\over2}$, 
and $\theta_{\rm P}=\arccos(z_{\rm P}/R_{\rm P})$ 
[i.e.\ $({\hat R}_{\rm P}, \theta_{\rm P}, \varphi_{\rm P})$ 
comprise the spherical coordinates of P in units of the light-cylinder radius $c/\omega$].

The expression in Eq.~\ref{eqnb} yields the electric-field vector of
the wave that propagates from the source to the detector directly.
The corresponding expression for the electric-field vector of the
reflected wave is given by a similar expression in which the
coordinates $(R_{\rm P}, \theta_{\rm P},\varphi_{\rm P})$ of P are
replaced by the coordinates $(R_{\rm r}, \theta_{\rm P'},\varphi_{\rm
P'})$ of the image P${}^\prime$ of P and the integral is multiplied by
one of the Fresnel coefficients $q_\parallel$ or $q_\perp$ (see the
preceding subsections):
\begin{equation}
{\bf E}_{\rm r}(f_\pm)\propto\Re\Big\{q_{\parallel,\perp}\exp(-{\rm i}n_\pm\omega t_{\rm P})      
\int_{-\pi/36}^{\pi/36} d\varphi\, \sin(\varphi-\varphi_{\rm P'})
\exp[{\rm i}(n_\pm{\hat R_{\rm r}}\pm m\varphi)]/{\hat R_{\rm r}}\Big\}{\hat{\bf e}}_{\rm r},
\label{eqnbr}
\end{equation}
where
\begin{equation}
{\hat R}_{\rm r}\equiv[{{\hat R}_{\rm P'}}^2+{\hat r}^2
-2{\hat r}{\hat R}_{\rm P'}\sin\theta_{\rm P'}\cos(\varphi-\varphi_{\rm P'})]^{1\over2}.
\end{equation}
The reflected radiation is polarized along 
\begin{equation}
{\hat{\bf e}}_{\rm rw}={\hat{\bf e}}_{\varphi_{\rm P}}
\end{equation}
if ${\bf E}(f_\pm)$ is normal to the plane of incidence, and along 
\begin{equation}
{\hat{\bf e}}_{\rm rw}=2({\hat{\bf e}}_{\rm g}\cdot{\hat{\bf e}}_{\varphi_{\rm P'}}){\hat{\bf e}}_{\rm g}
-{\hat{\bf e}}_{\varphi_{\rm P'}}
\label{pol}
\end{equation}
if ${\bf E}(f_\pm)$ is parallel to the plane of incidence, where
${\hat{\bf e}}_g$ is the upward-pointing unit vector perpendicular to
the ground (see Eqs.~\ref{refla} and \ref{reflb}).  The coordinates
$(R_{\rm P'}, \theta_{\rm P'},\varphi_{\rm P'})$ have the values
derived in Eq.~\ref{tfmnfour} when the array is positioned as in
Fig.~\ref{fig3}(a), and those derived in Eq.~\ref{tfmnfive} when the
array is positioned as in Fig.~\ref{fig3}(b). The values of
$q_\parallel$ and $q_\perp$ appear in Eqs.~\ref{qperp} and
\ref{qparallel}.

Next, we need to find the power that is received at P by an observer
whose detector only picks up radiation of a given polarization, {\em e.g.}\
a polarization along the unit vector ${\hat{\bf p}}$.  This can be
done by adding the components ${\hat{\bf p}}\cdot{\bf E}(f_\pm)$ and
${\hat{\bf p}}\cdot{{\bf E}_{\rm r}}(f_\pm)$ of the electric-field
vectors of the direct and the reflected waves and averaging the square
of the resulting expression over the observation time $t_{\rm P}$.
The result is
\begin{eqnarray}
{\cal P}_{\parallel,\perp}(f_\pm)\propto\bigg\vert&{\hat {\bf p}}\cdot{\hat{\bf e}}_{\varphi_{\rm P}}\int_{-\pi/36}^{\pi/36} d\varphi\, \sin(\varphi-\varphi_{\rm P})
\exp[{\rm i}(n_\pm{\hat R}\pm m\varphi)]/{\hat R}\cr
&+q_{\parallel,\perp}{\hat {\bf p}}\cdot{\hat{\bf e}}_{\rm rw}\int_{-\pi/36}^{\pi/36} d\varphi\, \sin(\varphi-\varphi_{\rm P'})
\exp[{\rm i}(n_\pm{\hat R}_{\rm r}\pm m\varphi)]/{\hat R}_{\rm r}\bigg\vert^2,
\label{power}
\end{eqnarray}
a general result that applies to both frequencies, irrespective of
whether the radiation is polarized parallel or perpendicular to the
plane of incidence.

When the array is positioned as in Fig.~\ref{fig3}(b), the generated
radiation is polarized horizontally, {\em i.e.}\ in a direction normal
to the plane of incidence.  In this case, ${\hat {\bf
p}}\cdot{\hat{\bf e}}_{\varphi_{\rm P}}$ and ${\hat {\bf
p}}\cdot{\hat{\bf e}}_{\rm rw}$ in the above equation are equal and so
can be factored out and absorbed in the constant of proportionality.
But when the plane of the array is vertical, as in Fig.~\ref{fig3}(a),
the electric-field vectors of the generated and reflected waves,
though both in the plane of incidence, are no longer parallel to each
other.  In such experiments, we have used a receiving antenna whose
polarization is vertical, {\em i.e.}\ for which ${\hat{\bf p}}={\hat{\bf
e}}_{\rm g}$.  According to Eqs.~\ref{normal} and \ref{pol},
therefore, the factors in question have the values ${\hat{\bf
p}}\cdot{\hat{\bf e}}_{\varphi_{\rm P}}=-\sin(\varphi_{\rm P}+\phi_0)$
and ${\hat{\bf p}}\cdot{\hat{\bf e}}_{\rm rw}=-\sin(\varphi_{\rm
P'}+\phi_0)$ in these experiments.

Note that the emission is beamed when the phase of the rapidly
oscillating exponential in the integrand of Eq.~\ref{eqnb} has a
stationary point within the domain of integration, {\em i.e.}\ when
\begin{equation}
{\partial\over{\partial\varphi}}[n_\pm{\hat R}\pm m(\varphi-\varphi_{\rm P})]
\end{equation}
vanishes for a value of $\varphi$ in $-\pi/36<\varphi<\pi/36$.  This
stationary point would be degenerate, and so the beaming would be
tighter, if
\begin{equation}
{\partial^2\over{\partial\varphi^2}}[n_\pm{\hat R}\pm m(\varphi-\varphi_{\rm P})]
\end{equation}
vanishes at the same time.  The above two conditions are satisfied
simultaneously when
\begin{equation}
{\hat r}=\frac{\omega r}{c}=
{m\over{\vert n_\pm\vert}}\Big[{{1-m^2/(n_\pm{\hat R}_{\rm P})^2}
\over {\sin^2\theta_{\rm P}-m^2/(n_\pm{\hat R}_{\rm P})^2}}\Big]^{1\over2},
\label{eqnf}
\end{equation}
and so the phase in question is stationary at $\varphi=\varphi_{\rm
P}+2\pi-\arccos[m^2/({n_\pm}^2{\hat r}{\hat r}_{\rm P})]$
(cf.~\cite{HAAAJS1,HAAAJS2}).  The parameters in our experiments are
chosen such that the value of $r$ implied by Eq.~\ref{eqnf}, for an
observer in the far field, lies within the radial thickness $10.00$ m
$<r<$ $10.05$ m of the array.

\subsection*{Adjustable parameters: effects of the
array orientation and surface corrugations on the modeling of the
beaming data} 
There are no truly adjustable parameters in the analysis
of the beaming data (Figs.~\ref{expfiga}-\ref{phibeamshape}) in this
paper, especially when $R~\gtsim ~300$~m. The speed of the source is
constrained by the setting of the experimental machine and the
refractive index of the ground is determined by anciliary experiments
with the dipole aerial (see Fig.~\ref{intscheme}).  The measured
heights $h_{\rm A}$ and $h_{\rm D}$ and distance $R$ are also used as
input parameters.  The remaining uncertainties come from the
difficulties involved with modeling the reflection of curved wave
fronts and small inaccuracies in the angular coordinates. We treat the
latter first.

In modeling the experimental data of
Figs.~\ref{expfiga}-\ref{phibeamshape}, variations in the parameters
$\vartheta_0$ (see Eq.~\ref{norman}) and $\phi_0$ (see
Eq.~\ref{normal}), the angles marking the orientation of the array,
have a significant effect on the relative phases of the rays traveling
on the direct and reflected paths (Fig.~\ref{fig12}).  Ideally, with
flat ground ({\it i.e.}\ the runway) and precision measurements of the
orientation of the array, these angles would be completely constrained
by the measured experimental coordinates $\phi_{\rm V}$ and
$\vartheta_{\rm H}$ respectively, plus the known $h_{\rm A}$, $h_{\rm
D}$ and $R$. However, there are two problems with this approach:
\begin{enumerate}
\item
The measurement of the angular coordinate desribing
the tilting of the array is
subject to relatively large fractional 
errors, $\sim \pm 0.5^{\circ}$;
\item
The runway surface used for the measurement has distinct corrugations
corresponding to fractions of a degree; over long distances, these can
result in non-negligible shifts in the apparent position of the
reflected ray.
\end{enumerate}
These uncertainties were encompassed by making $\vartheta_0$ or
$\phi_0$ (as appropriate) somewhat adjustable; adjustments of less
than $\sim 1^{\circ}$ from the expected value ({\it i.e.}\ of the
order of the uncertainty in the experimental angles) were in general
sufficient to obtain very close agreement between model and data.

A second cause of uncertainty is the use of Fresnel coefficients in
dealing with reflection from the ground; Fresnel coefficients are of
course derived for plane wave fronts~\cite{bleaney}.  As is well known
in the modeling of radar beams~\cite{Stratton}, wave fronts close to a
source have significant curvature, leading to an apparent loss of the
reflected radiation due to ``phase smearing''. In the present case,
the effect becomes noticeable for $R~\ltsim ~300$~m.

A detailed treatment of curved wave fronts is very complex, and beyond
the scope of this paper. However, a number of empirical approximations
may be used~\cite{Stratton}.  Following these precedents, the
reflection coefficient in simulations of the beaming involving
distances $R~\ltsim ~300$~m was scaled by a factor $q_0$, representing
the loss of signal on reflection due to the curved wave fronts.

\end{document}